\documentclass[11pt]{article}
\pdfoutput=1

\usepackage{amsmath}
\usepackage{amssymb}
\usepackage{amsfonts}
\usepackage{mathrsfs}

\usepackage{mathrsfs}
\usepackage{fullpage}
\usepackage{setspace}
\usepackage{bm}
\usepackage{bbm}

\usepackage[normalem]{ulem}
\usepackage{enumerate}
\usepackage{yfonts}

\usepackage{psfrag}

\usepackage{graphicx}
\usepackage{cancel}
\usepackage{slashed}

\usepackage[font=small,labelfont=bf]{caption}
\usepackage{subcaption}

\usepackage[colorlinks=true]{hyperref} 
\hypersetup{
    bookmarks=true,         
    unicode=false,          
    pdftoolbar=true,        
    pdfmenubar=true,        
    pdffitwindow=false,     
    pdfstartview={FitH},    
    pdftitle={Displacement memory effect near the horizon of black holes},    
    pdfauthor={Srijit Bhattacharjee, Shailesh Kumar, Arpan Bhattacharyya},     
    pdfnewwindow=true,      
    colorlinks=true,       
    linkcolor=blue,          
    citecolor=red,        
    filecolor=magenta,      
    urlcolor=cyan           
} 

\begin{document}

\numberwithin{equation}{section}
{
\begin{titlepage}
\begin{center}

\hfill \\
\hfill \\
\vskip 0.75in

{\Large \bf Displacement memory effect near the horizon of black holes}\\

\vskip 0.4in

{\large
Srijit Bhattacharjee${}^{a}$\footnote{srijuster@gmail.com}, Shailesh Kumar${}^{a}$\footnote{shaileshkumar.1770@gmail.com} and Arpan Bhattacharyya${}^{b}$\footnote{abhattacharyya@iitgn.ac.in}}

\vskip 0.3in

{\it ${}^{a}$ Indian Institute of Information Technology (IIIT), Allahabad,\\
Devghat, Jhalwa, Uttar Pradesh-211015, India} \vskip .5mm
{\it ${}^{b}$Indian Institute of Technology, Gandhinagar, Gujarat-382355, India}

\vskip.5mm

\end{center}

\vskip 0.35in

\begin{center} {\bf ABSTRACT } \end{center}
We study the displacement memory effect and its connection with the extended-BMS symmetries near the horizon of black holes. We show there is a permanent shift in the geodesic deviation vector relating two nearby timelike geodesics placed close to the horizon of black holes, upon the passage of gravitational waves. We also relate this {\it memory} effect with the asymptotic symmetries near the horizon of asymptotic black hole spacetimes. The shift of the relative position of the detectors is shown to be induced by a combination of BMS generators near the horizon. The displacement memory effect near the horizon possesses similarities to the same obtained in the far region.  
\vfill


\end{titlepage}
}

\newpage
\tableofcontents

\section{Introduction}
The direct detection of gravitational wave (GW) \cite{PhysRevLett.116.061102, PhysRevLett.125.101102} has enabled researchers to look for various aspects of black hole spacetimes; ``Gravitational memory effect'' \cite{Zeldovich:1974gvh, Braginsky-Thorne, Braginsky, PhysRevLett.67.1486, Bieri:2013ada, Bieri:2015yia, PhysRevD.80.024002} is one of such physical effects which manifests the permanent relative separation or displacement between the test masses, initially held at relative rest, upon interacting with GWs. This is nothing but the conventional displacement memory effect. This fascinating field recently has gained considerable attention in the GW physics due to its possible detection in the advanced LIGO or LISA detectors \cite{Islo:2019qht,PhysRevLett.117.061102,PhysRevD.101.083026}. On the other hand, from theoretical perspectives, it has brought interests of many due to its connection with Bondi-Metzner-Sachs (BMS) symmetries \cite{Bondi:1962px} and soft graviton theorem \cite{Strominger:2014pwa, Strominger:2017zoo, PhysRevLett.116.231301, Hawking:2016sgy, Zhang:2017geq}. It has been shown that asymptotic BMS symmetries can be studied in the context of memory effect \cite{Strominger:2017zoo}. Further, the fundamental relation between GW-memory and soft graviton theorem has been established by Strominger and Zhiboedov \cite{Strominger:2014pwa, Pasterski:2015tva}. Such connections with memory effect have offered us an intriguing opportunity to delve into the low energy quantum gravity picture \cite{Strominger:2014pwa, Strominger:2017zoo, PhysRevLett.116.231301, Hawking:2016sgy}. Recently, Hawking, Perry and Strominger conjectured that the charges corresponding to BMS symmetries would help to retrieve the information in the Hawking information paradox \cite{PhysRevLett.116.231301, hawking2015information}. Therefore, in this respect, BMS symmetries for near-horizon gravitational memory might play a crucial role in understanding the information paradox.

It is known that the supertranslation-like BMS symmetries and the extension of that \cite{PhysRevLett.105.111103, Barnich:2011ct} can also be recovered at the horizon of black hole spacetimes by studying the asymptotic symmetries preserving the near-horizon structure of three and four-dimensional spacetimes \cite{PhysRevLett.116.091101, Donnay2016, PhysRevD.98.124016}. Non-extremal black holes in four-dimensional general relativity exhibit an infinite-dimensional symmetry in their near-horizon region. One can show, by employing suitable boundary conditions that the algebra of asymptotic Killing vectors is infinite-dimensional. It contains two sets of supertranslations and two mutually commuting copies of the Virasoro algebra \cite{PhysRevLett.116.091101, Donnay2016, PhysRevD.98.124016}. For stationary backgrounds, one of the sets of Virasoro algebras freezes off. It should also be noted that the BMS symmetries, including superrotations like transformations, can also be recovered near the black hole horizon demanding the invariance of the induced metric on a null hypersurface situated at the horizon and serving as a boundary of two black hole spacetimes \cite{Blau:2015nee, Bhattacharjee}. The memory effect related to these horizon shells containing impulsive gravitational waves and for plane wave spacetimes is recently being studied in \cite{OLoughlin:2018ebk, PhysRevD.100.084010,PhysRevD.102.044041, p7}. Memory effect near the black hole horizon in terms of BMS charge algebra has been studied in \cite{PhysRevD.98.124016}. Interesting studies on the memory effect related to Rindler spacetimes can be found in \cite{Kolekar:2017yoi, Compere:2019rof}.

Motivated by the Hawking, Perry, Strominger's proposal \cite{Hawking:2016sgy} and the detection prospect, we provide a detailed analysis of computing the near-horizon displacement memory effect and find its connection with the BMS symmetries recovered near the horizon. Our particular interest is to see the effect of near-horizon BMS symmetries \cite{PhysRevD.98.124016} on the test detectors placed near the horizon by computing their permanent shift upon passage of gravitational waves. To compute the displacement memory effect, one starts with the ``geodesic deviation equation'' (GDE) which measures the deviation or displacement between two nearby timelike geodesics. We focus on estimating the memory effect for the test detectors stationed near the black hole horizon in three and four dimensions. Although the present study can at best be regarded as just a model at the moment, however in future this may provide a window to study the near-horizon memory effect with an advanced detection mechanism. 

In order to understand the detection framework, consider two nearby timelike geodesics or test masses of GW detectors positioned near the horizon of a black hole spacetime with a tangent vector $T^{\mu}$ and deviation or separation vector $S^{\mu}$. The deviation vector $S^{\mu}$ between adjacent geodesics evolves according to GDE, given by
\begin{align}
    \frac{D^{2}}{d\tau^{2}}S^{\mu} = -R^{\mu}{}_{\alpha\beta\gamma}T^{\alpha}T^{\gamma}S^{\beta},\label{deviation}
\end{align}
where $\tau$ is the proper time, $S^{\mu}$ denotes the spatial separation between two nearby timelike geodesics, and $R^{\mu}{}_{\alpha\beta\gamma}$ is the Riemann tensor. We shall adopt Greek letters for spacetime index ($\mu = 0, 1, 2, 3$), Latin lower case letters for hypersurface index ($a = 1, 2, 3$) and Latin upper case letters for the spatial coordinates
on the hypersurface, i.e. coordinates for the co-dimension one surface ($A = 2, 3$). The GDE, in a particular coordinate system, is given by \cite{Bieri:2015yia}
\begin{align}
    \Ddot{S}^{i} = -R_{titj} S^{j}. \label{BGDE}
\end{align}
In weak field slow motion approximation, the Riemann tensor depends on the time derivatives of the Quadrupole moment tensor ($Q_{ij}$), $R_{titj} = -\frac{1}{r}\mathcal{P}[\ddddot{Q}_{ij}]$.
Where $\mathcal{P}$ function gives ‘projected orthogonally to the radial direction and trace-free' part of the tensor on which it acts. The interaction between the detectors and GW should induce a permanent change in the spatial separation of the setup. This can be seen by integrating the Eq. (\ref{BGDE}) twice.
\begin{align}
    \Delta S^{i} = \frac{1}{r}S^{j}\mathcal{P}[\Delta\ddot{Q}_{ij}], \label{BiSep}
\end{align}
where `$\Delta$' denotes the difference between the separation vector before and after the passage of GW. This spatial separation between the detectors, to the linear order of $\frac{1}{r}$, depicts the conventional memory effect. It should be noted here that the Eq. (\ref{BiSep}) is only valid for the linear approximation to general relativity and for slowly moving sources. In this report, we focus on the near vicinity of the horizon, and should not consider the weak field slow-motion approximation. Nonetheless, we can still obtain an analogous effect given by a change in the metric parameter in the right-hand side of Eq.  (\ref{BiSep}) in a suitable coordinate system. Near the horizon, the jump in the separation is governed by the change in the metric parameters instead of Quadrupole moment tensor,

\[\Delta S^i\sim \rho \Delta B_{ij}S^j,\] where $B$ denotes a function of the coordinates carried by the metric, and $\rho$ denotes the radial coordinate. Further, we show that the change in the separation of deviation vector can be realized by a combination of supertranslation and superrotation revealing the BMS memory effect near the horizon of black hole spacetimes. 

 Let us briefly discuss how the paper has been organized. We start by reviewing some basic details in section (\ref{2D}). In section (\ref{3D}), we provide a detailed analysis of displacement memory effect for three-dimensional near-horizon extreme and non-extreme black holes. The similar analysis, we have provided for the four-dimensional black holes in section (\ref{4D}). This section contains discussion of memory effect for the full asymptotic metric as well as a less generic metric. We also establish the connection between the memory effect and BMS symmetries.  In the discussions, we sum up our findings and provide some outlooks.
\section{Memory effect at null infinity in flat spacetimes} \label{2D}
The gravitational memory effect for the asymptotically flat spacetimes near future null infinity has been a well-studied area. The inertial detectors positioned near the future null infinity get permanently displaced after the passage of GW. This is regarded as the conventional gravitational memory effect \cite{Zeldovich:1974gvh,Braginsky-Thorne,PhysRevLett.67.1486,Strominger:2014pwa,Strominger:2017zoo,Strominger2017,PhysRevD.89.064008,PhysRevD.45.520,PhysRevD.90.044060}. It has also been established that there is a direct relation between this displacement and the BMS symmetries \cite{Strominger:2014pwa,Strominger2017}. Recently, it has been explicitly shown in \cite{Strominger:2017zoo} how the displacement memory effect at the far region is related to a supertranslation. To see this, consider the general form of the asymptotically flat metric
\begin{align}
ds^{2} =& -du^{2}-2du dr+2r^{2}\gamma_{z\bar{z}}dz d\bar{z}+\frac{2m_{B}}{r}du^{2}+D^{z}C_{zz}du dz+ D^{\bar{z}}C_{\bar{z}\bar{z}}du d\bar{z}+ \nonumber \\ 
& rC_{zz}dz^{2}+rC_{\bar{z}\bar{z}}d\bar{z}^{2}+\frac{1}{r}\Big(\frac{4}{3}(N_{z}+u\partial_{z}m_{B})-\frac{1}{4}\partial_{z}(C_{zz}C^{zz})\Big) du dz+c.c.+..., \label{asymfd}
\end{align}
where $m_{B}$, $N_{z}$ in general depend on ($u,z,\bar{z}$); and they are also called as ``Bondi mass'' and ``angular momentum aspect'' respectively. Whereas $C_{zz}(u,z,\bar{z})$ describes GWs. It is, in fact, the free data available near asymptotic null infinity. It is related to the ``Bondi news tensor'' ($N_{zz}$), written as $N_{zz}=\partial_{u}C_{zz}$. We can see that the first three terms in the metric (\ref{asymfd}) represent the flat Minkowski spacetime and other terms represent the leading order correction to the metric. ``...'' tells about the subleading terms at large $r$. For large $r$ fall-offs, the metric (\ref{asymfd}) components are
\begin{align}
g_{uu} =& -1+\mathcal{O}(\frac{1}{r}) \hspace{3mm} ; \hspace{3mm} g_{ur} = -1+\mathcal{O}(\frac{1}{r}) \hspace{3mm} ; \hspace{3mm} g_{uz} = \mathcal{O}(1) \nonumber \\
g_{zz} =& \mathcal{O}(r) \hspace{3mm} ; \hspace{3mm} g_{z\bar{z}} = r^{2}\gamma_{z\bar{z}}+\mathcal{O}(r) \hspace{3mm} ; \hspace{3mm} g_{rr} = g_{rz} = 0 \label{asymfd1}
\end{align} 
The geodesic deviation equation for the metric (\ref{asymfd}) takes the following form
\begin{align}
r^{2}\gamma_{z\bar{z}}\partial^{2}_{u}S^{\bar{z}} = -R_{uzuz}S^{z}, \label{asymr}
\end{align}
where $R_{uzuz} = -\frac{r}{2}\partial^{2}_{u}C_{zz}$. Therefore, the change in the displacement is given by
\begin{align}
\Delta S^{\bar{z}} = \frac{\gamma^{z\bar{z}}}{2r}\Delta C_{zz}S^{z}. \label{asymdev}
\end{align} 
From the $uu$-component of the Einstein field equation (EFE), and for a stress tensor with shockwave profile of the form
\begin{align}
T_{uu}(u,z,\bar{z})=\mu\delta(u-u_{rad})\frac{\delta^{2}(z-z_{rad})}{\gamma_{z\bar{z}}}, \label{str4dasmp}    
\end{align}
the leading order change in the $C_{zz}$ is given by
\begin{align}
\Delta C_{zz}(z,\bar{z}) =& C_{zz}(z,\bar{z})\vert_{u=u_{f}}-C_{zz}(z,\bar{z})\vert_{u=u_{0}} \nonumber \\
\Delta C_{zz}(z,\bar{z}) =& 2\mu D^{2}_{z}G(z,\bar{z};z_{rad},\bar{z}_{rad})-\frac{\mu}{2\pi}\int d^{2}z'\gamma_{z'\bar{z}'}D^{2}_{z}G(z,\bar{z};z',\bar{z}'),
 \label{asydel4d}
\end{align}
where $G(z,\bar{z};z_{rad},\bar{z}_{rad})$ is the Green's function can be found in \cite{Strominger2017}. The above expression contributes to the deviation equation to obtain the memory effect at the far region. 

Further, to relate the memory effect with BMS symmetry, consider the supertranslation of type $u\rightarrow u+f(z,\bar{z})$. One can find a $f(z,\bar{z})$ which would give rise the same change in $C_{zz}$ of (\ref{asydel4d}). For this, taking Lie derivative of $C_{zz}$ along the supertranslation parameter $f$ we get,
\begin{align}
\mathcal{L}_{f}C_{zz} = f N_{zz}-2D^{2}_{z}f. \label{lie4d1}
\end{align}
Eq. (\ref{lie4d1}) is directly related to change in the $C_{zz}$. If before and after the passage of GW the $N_{zz}$ is zero, then one obtains
\begin{align}
\Delta C_{zz} = -\mathcal{L}_{f}C_{zz} = 2D^{2}_{z}f. \label{lie4d2} 
\end{align}
Now one can choose
\begin{align}
f(z,\bar{z}) = \mu G(z,\bar{z};z_{rad},\bar{z}_{rad})-\frac{\mu}{4\pi}\int d^{2}z'\gamma_{z'\bar{z}'}D^{2}_{z}G(z,\bar{z};z',\bar{z}').
\end{align}
This choice of $f$ produces the same change in the $\Delta C_{zz}$ as appears in (\ref{asydel4d}). This establishes the connection between memory effect and BMS symmetries near the future null infinity for asymptotic flat spacetimes. Next we provide a similar study and connection with BMS symmetries for the near-horizon asymptotic form of black holes.

\section{Near-Horizon Memory for Three-Dimensional Black Holes}\label{3D}
Three-dimensional gravity theories sometimes give us a good hint of what should be expected in the four-dimensional scenario. In many cases, analysing three-dimensional models becomes easy. Therefore, to start with, we analyse the three-dimensional near-horizon metric for anti-de Sitter (AdS) spacetime, and compute the displacement memory effect. We are interested in measuring the change in the relative displacement or deviation vector between two nearby timelike geodesics or test detectors, induced via the interaction with GWs generated from the black hole spacetime. Let us consider the near-horizon metric for a three-dimensional black hole in Gaussian null coordinates \cite{PhysRevLett.116.091101,Donnay2016,PhysRevD.98.124016}.
\begin{align}
ds^{2}=\xi dv^{2}+2kdvd\rho +2h dvd\phi +R^{2}d\phi^{2}, \label{1}
\end{align}
where $v$ is the temporal coordinate, $\rho \geq 0$ represents the radial distance to the horizon and $\phi$ is the angular coordinate of period $2\pi$. Functions $\xi$, $k$, $h$, and $R$ are expected to comply with the following fall-off conditions near the horizon, which is at $\rho = 0$:
\begin{align}
\xi =& -2\kappa\rho +\mathcal{O}(\rho^{2}) \hspace{3mm} ; \hspace{8mm} k = 1+\mathcal{O}(\rho^{2}) \nonumber\\
h =& \theta (\phi)\rho + \mathcal{O}(\rho^{2}) \hspace{5mm} ; \hspace{8mm} R^{2} = \gamma(\phi)^{2}+\lambda(v,\phi)\rho + \mathcal{O}(\rho^{2}) \label{2}
\end{align}
where $\kappa(v,\phi)$, $\theta(\phi)$, $\gamma(\phi)$ and $\lambda(v,\phi)$ are arbitrary functions of the coordinates. 
The metric components $g_{\rho\phi}$ and $g_{\rho\rho}$ decay rapidly as $\mathcal{O}(\rho^{2})$. 

The asymptotic boundary conditions are being preserved by asymptotic Killing vectors, given by \cite{Donnay2016}
\begin{align}
Z^{v} =& f(v,\phi) \nonumber\\
Z^{\rho} =& -\partial_{v}f \rho + \partial_{\phi}f \frac{\theta}{2\gamma^{2}}\rho^{2}+\mathcal{O}(\rho^{3}) \label{killing3D} \\
Z^{\phi} =& Y(\phi)-\partial_{\phi}f\frac{\rho}{\gamma^{2}}+\partial_{\phi}f\frac{\lambda}{2\gamma^{4}}\rho^{2}+\mathcal{O}(\rho^{3}), \nonumber 
\end{align}
where $f(v,\phi)$ and $Y(\phi)$ are arbitrary functions, and prime denotes the derivative with respect to $\phi$. In addition to this, $Z^{\rho}$ may contain an $\mathcal{O}(1)$ term $\tilde{z}(v,\phi)$. However, we set such terms to be zero provided $Z=0, \partial_{v}Y=0$.  Under such transformations, the arbitrary functions transform along the killing direction as follows
\begin{align}
\delta_{Z}\kappa =& Y\partial_{\phi}\kappa + \partial_{v}(\kappa f)+\partial^{2}_{v}f \nonumber\\
\delta_{Z}\gamma =&\partial_{\phi}(Y\gamma)+f\partial_{v}\gamma \nonumber\\
\delta_{Z}\theta =& \partial_{\phi}(Y\theta)+f\partial_{v}\theta-2\kappa\partial_{\phi}f-2\partial_{v}\partial_{\phi}f+2\partial_{\phi}f\frac{\partial_{v}\gamma}{\gamma} \label{trans3Dn1}\\
\delta_{Z}\lambda =& Y\partial_{\phi}\lambda +2\lambda\partial_{\phi}Y+2\theta\partial_{\phi}f-2\partial^{2}_{\phi}f+2\partial_{\phi}f\frac{\partial_{\phi}\gamma}{\gamma}+f\partial_{v}\lambda -\lambda\partial_{v}f . \nonumber 
\end{align}
Introducing the modified Lie bracket
\begin{align}
[Z_{1},Z_{2}] = \mathcal{L}_{Z_{1}}Z_{2}-\delta_{Z_{1}}Z_{z}+\delta_{Z_{2}}Z_{1}, \label{mbrcs3D}
\end{align}
One gets
\begin{align}
[Z(T_{1},X_{1},Y_{1}),Z(T_{2},X_{2},Y_{2})] = Z(T_{12},X_{12},Y_{12}), \label{alg3D}
\end{align}
where
\begin{align}
T_{12} =& Y_{1}\partial_{\phi}T_{2}-Y_{2}\partial_{\phi}T_{1} \nonumber \\  
X_{12} =& Y_{1}\partial_{\phi}X_{2}-Y_{2}\partial_{\phi}X_{1}-\kappa (T_{1}X_{2}-T_{2}X_{1})   \\
Y_{12} =& Y_{1}\partial_{\phi}Y_{2}-Y_{2}\partial_{\phi}Y_{1}. \nonumber
\end{align}
Here we considered fixed temperature configuration i.e. fixed $\kappa$ for which 
\begin{align}
    f(v,\phi) = T(\phi)+e^{-\kappa v} X(\phi), \label{fxdT3D}
\end{align}
where $T(\phi)$ and $X(\phi)$ are two sets of supertranslation generators \cite{PhysRevLett.116.091101, Donnay2016}. 

One can further establish the algebra between generators $T_{n}$, $X_{n}$ and $Y_{n}$ by defining Fourier modes, $T_{n}=Z(e^{in\phi},0,0)$, $X_{n}=Z(0,e^{in\phi},0)$ and $Y_{n}=Z(0,0,e^{in\phi})$, given by
\begin{align}
    i[Y_{m},Y_{n}] =& (m-n)Y_{m+n} \hspace{3mm} ; \hspace{3mm}
    i[Y_{m},T_{n}] = -nT_{m+n} \nonumber \\
    i[Y_{m},X_{n}] =& -nX_{m+n} \hspace{9mm} ; \hspace{3mm} i[T_{m},X_{n}] = -\kappa X_{m+n}  \label{alg3D1}
\end{align}
The generator $T_{n}$ appears as a supertranslation from the symmetry of type $v\rightarrow v+T(\phi)$ and $Y_{n}$ causes superrotations. The conserved charges corresponding to the symmetries have also been found in \cite{PhysRevLett.116.091101,Donnay2016,PhysRevD.98.124016} and form a representation of (\ref{alg3D1}). The zero modes of the generators produce the Bekenstein-Hawking entropy of stationary black holes \cite{PhysRevLett.116.091101,Donnay2016}. 

In the extremal scenario for which $\kappa=0$, the charge associated with the zero mode of $X$ is the one that gives the entropy \cite{Donnay2016}.

\subsection{Geodesic Deviation Equation and Memory Effect}
Now, let us find the geodesic deviation equation for the metric (\ref{1}). As mentioned earlier that in general $\kappa$ depends on ($v,\phi$), the deviation equation takes the following form\footnote{Here, we have considered terms leading order in $\rho$ for the Riemann tensor.}
\begin{align}
\gamma^{2}(\partial_{v}^{2}S^{\phi}-\kappa(v,\phi)\partial_{v}S^{\phi}) = -R_{\phi v\phi v} S^{\phi},\label{4}
\end{align}
where the Riemann tensor component is given by
\begin{align}
R_{\phi v\phi v} =& \rho\Big(\frac{1}{2}\kappa(v,\phi) \theta (\phi)^{2}+\kappa(v,\phi)^{2}\lambda(v,\phi)+\frac{\kappa(v,\phi)\theta(\phi)\gamma(\phi)^{'}}{\gamma(\phi)}-\kappa(v,\phi)\theta(\phi)^{'}-\frac{\gamma(\phi)^{'}}{\gamma(\phi)}\partial_{\phi}\kappa(v,\phi)+ \nonumber \\ 
& \partial_{\phi}^{2}\kappa(v,\phi)-\frac{1}{2}\lambda(v,\phi)\partial_{v}\kappa(v,\phi)+\frac{1}{2}\kappa(v,\phi)\partial_{v}\lambda(v,\phi)-\frac{1}{2}\partial^{2}_{v}\lambda(v,\phi)\Big)+\mathcal{O}(\rho^{2}).\label{55}
\end{align}
Here prime denotes the derivative of the function with respect to $\phi$ coordinate. However, as our motivation is to find the memory near black holes, we consider only fixed temperature configurations i.e. cases where $\kappa$ is constant. We also set $\partial_v\gamma=0$. Under these assumptions, the Riemann tensor component becomes
\begin{align}
R_{\phi v\phi v} = \rho\Big(\frac{1}{2}\kappa \theta (\phi)^{2}+\kappa^{2}\lambda(v,\phi)+\frac{\kappa\theta(\phi)\gamma(\phi)^{'}}{\gamma(\phi)}-\kappa\theta(\phi)^{'}+\frac{1}{2}\kappa\partial_{v}\lambda(v,\phi)-\frac{1}{2}\partial^{2}_{v}\lambda(v,\phi)\Big)+\mathcal{O}(\rho^{2}),\label{5}
\end{align}
where $\kappa = \frac{1}{l^{2}r_{+}}(r_{+}^{2}-r_{-}^{2})$; $r_{+}$ and $r_{-}$ denoting the outer and inner horizons respectively. $l$ is the length scale of AdS spacetime. 

\subsubsection{Extremal Case}
Above equations (Eq. \ref{4}, \ref{5}) considerably simplify in the case when $\kappa=0$.  In this case,  the $g_{vv}$ component of the asymptotic metric becomes, $g_{vv} = L(v,\phi)\rho^{2} + \mathcal{O}(\rho^{3})$ \cite{PhysRevD.98.124016}, and therefore does not contribute to the leading order. Now, for this extremal form of the metric, the geodesic deviation equation (\ref{deviation}) takes the following form
\begin{align}
\gamma(\phi)^{2}\partial^{2}_{v}S^{\phi}_{E} = \frac{\rho}{2}\partial^{2}_{v}\lambda S^{\phi}_{E},
\end{align}
where $S^{\phi}_{E}$ depicts the $\phi$ component of the deviation vector for extremal case. 

\begin{figure}[h!]
	\centering
	\includegraphics[width=75mm]{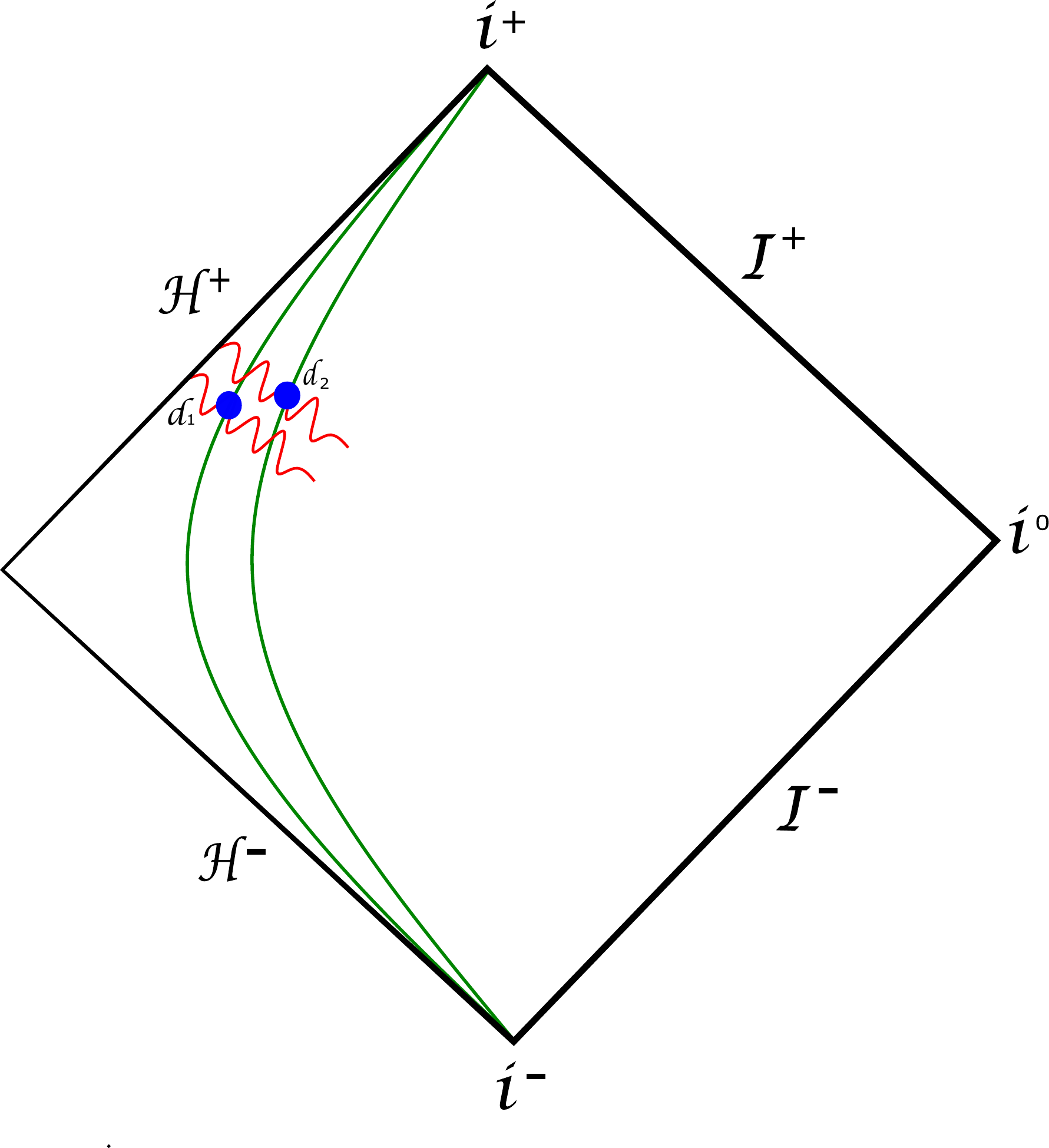}
	\caption{A schematic diagram of the displacement memory effect near the horizon of a black hole. The pulses of gravitational radiation pass the detectors and generate a permanent change in the relative position. }
	\label{FIGURE}
\end{figure}

Now we can measure the change in the deviation vector which can be written as
\begin{align}
\Delta S^{\phi}_{E} =\frac{\rho}{2\gamma(\phi)^{2}}(\Delta \lambda) S^{\phi}_{E} + \mathcal{O}(\rho^{2}).\label{gd1}
\end{align}
Here we observe that our goal is to find the change in $\lambda(v,\phi)$ in order to estimate the jump in the deviation vector. For this, we solve the $vv$-component of the Einstein field equation, given by
\begin{align}
-\frac{\rho}{2\gamma(\phi)^{2}}\partial^{2}_{v}\lambda = 8\pi T^{M}_{vv}. \label{f1}
\end{align}
Further, defining $T_{vv}=\lim_{\rho \to 0}8\pi\frac{T_{vv}^{M}}{\rho}$ and considering a shock-wave type profile for the stress tensor as 
\begin{align}
    T_{vv}(v,\phi) = s \delta(v-v_{0})g(\phi) ,\label{tvv}
\end{align} where $s$ is just a constant, the solution of the field equation becomes
\begin{align}
\lambda(v,\phi) = c_{1}(\phi)+\Big(c_{2}(\phi) -2s\gamma(\phi)^{2}g(\phi)\Big)v. \label{ful1}
\end{align}
To compute the change in $\lambda(v,\phi)$ i.e. $\Delta\lambda(\phi)$ for some initial and final $v$, we set the boundary conditions as following
\begin{align}
\lambda(v=v_i,\phi) = \lambda_{0}(\phi) \hspace{5mm} ; \hspace{5mm} \lambda(v=v_f,\phi) = \lambda_{f}(\phi).
\end{align}
Hence, the full solution can be written as
\begin{align}
\lambda(v,\phi) =C(\phi)+ F(\phi)v, \label{ld3d}
\end{align}
for some function $F(\phi)$ that contains the metric parameters only depending on $\phi$. Therefore, the change in $\lambda(v,\phi)$ can be written as
\begin{align}
\Delta\lambda = \lambda(v=v_{f},\phi)-\lambda(v=v_{i},\phi) = F(\phi)\Delta v, \label{lambda1}
\end{align}
where $\Delta v = v_{f}-v_{i}$. We plug this expression of $\Delta\lambda$ into (\ref{gd1}). This gives us the change in the  deviation vector $S^{\phi}_{E}$.
\begin{align}
\Delta S^{\phi}_{E} = \frac{\rho}{2\gamma(\phi)}F(\phi)\Delta v S^{\phi}_{E}+\mathcal{O}(\rho^{2}). \label{extreme3D}
\end{align}
This provides the change in the deviation vector between the test geodesics near the horizon for three-dimensional extreme black holes. This is the permanent change in the setup induced by the interaction with GWs and regarded as the displacement memory effect. Figure~(\ref{FIGURE}) provides an illustration of the situation we are examining. 

\subsubsection{Relation with BMS symmetry} \label{bms3d}

To see the connection between this memory effect and BMS symmetry, we consider the change in the parameter $\lambda$. For extreme case, the $v$ component of the Killing vector, generating two sets of supertranslations $T(\phi)$ and $X(\phi)$  becomes 
\begin{equation}
    f(v,\phi)=T(\phi)+v X(\phi).
\end{equation}
In general the $\lambda(v,\phi)$ can be written as $\lambda(v,\phi)=C(\phi)+v F(\phi)$, as shown in (\ref{ld3d}). If  $\Delta v$ is the difference in advanced-time before and after the passage  of GW, then
\begin{align}
\Delta \lambda= F(\phi)\Delta v . \label{dlmbda1}
\end{align}
This is an equivalent way of writing (\ref{lambda1}). Now, we consider the change in $\lambda(v,\phi)$ induced by asymptotic Killing vector $Z$ \cite{PhysRevD.98.124016},
\begin{align}
{\cal L}_{Z}\lambda =2 \lambda \partial_{\phi}Y+Y\partial_{\phi} \lambda + 2\theta \partial_{\phi} f-2 \partial_{\phi}^2 f +2\partial_{\phi}f\frac{\partial_{\phi}\gamma}{\gamma(\phi)}+f\partial_{v}\lambda -\lambda \partial_v f . \label{sup1}
\end{align}
Our goal is to produce Eq. (\ref{lambda1}) like change via ${\cal L}_{Z}\lambda$. Here, to make the variation of $\lambda(v,\phi)$ along the Killing direction $Z$ to be independent of $v$, we set coefficient of $v$ as zero. This will make sure the whole right hand side of the Eq. (\ref{sup1}) matches with the displacement in $\lambda$ derived in (\ref{lambda1}). We obtain the following condition
\begin{align}
    2F\partial_{\phi}Y+Y\partial_{\phi}F+2\theta\partial_{\phi} X-2\partial^{2}_{\phi}X+2\partial_{\phi}X\frac{\partial_{\phi}\gamma (\phi)}{\gamma(\phi)} = 0, \label{3d-cndn}
\end{align}
where $F$, $X$, $Y$ and $\theta$ are functions of $\phi$. We observe the condition of $\Delta \lambda(v,\phi)$ to be $v$ independent involves both $Y(\phi)$ and $X(\phi)$. The remaining terms in Eq. (\ref{sup1}) are functions of $\phi$ only and can be arranged in such a way that they become equal to $\Delta v F(\phi)$. To see the feasibility of such BMS transformation, we may consider only superrotation to be acting on the metric parameters, in that case Eq. (\ref{3d-cndn}) offers a solution $Y(\phi)=F^{-1/2}.$ On the other hand in the absence of superrotations the above equation provides a solution for $X(\phi)\propto \int d\phi e^{-\int A(\phi) d\phi}.$ Where $A(\phi)$ is a function of metric parameters $\theta$, $\gamma$ etc. In general one may also get a solution when both supertranslation and superrotation are present by choosing an appropriate $g(\phi)$ sitting in the expression of energy momentum tensor (\ref{tvv}). All these results confirm that the memory can be induced by BMS parameters superrotation $Y(\phi)$ and supertranslation $X(\phi)$.

\subsection{Non-Extremal Case}\label{nextml1}
For nonzero but constant $\kappa$ configuration, we again need to find the change in $\lambda(v,\phi)$. This can  obtained by solving the $vv$-component of the Einstein's field equation (to the leading order in $\rho$) 
\begin{align}
    -\frac{\rho}{2l^{2}\gamma^{3}}\Big(-4\kappa\gamma^{3}+2l^{2}\kappa\theta\gamma^{'}+l^{2}\gamma(\kappa\theta^{2}+2\kappa^{2}\lambda(v,\phi)-2\kappa\theta^{'}+3\kappa\partial_{v}\lambda(v,\phi)+\partial^{2}_{v}\lambda(v,\phi))\Big) = 8\pi T^{M}_{vv}. \label{6}
\end{align}
We shall again define $T_{vv}=\lim_{\rho \to 0}8\pi\frac{T^{M}_{vv}}{\rho}$ and consider the terms, other than $\lambda(v,\phi)$, as a function  $\kappa\, \mathcal{F(\phi)}$, defined by,
\begin{align}\mathcal{F} (\phi)=\rho\Big(\frac{2}{l^{2}}-\frac{1}{\gamma(\phi)^{2}}(\frac{\theta(\phi)\gamma(\phi)^{'}}{\gamma(\phi)^{2}}+\frac{\theta(\phi)^{2}}{2}-\theta(\phi)^{'}) \Big). \end{align}
Since we are interested to solve the field equation (\ref{6}) with respect to $v$ coordinate, $\phi$-dependent functions would be unaffected while solving the equation. Now Eq. (\ref{6}) becomes
\begin{align}
-\frac{1}{2\gamma(\phi)^{2}}\frac{\partial^{2}}{\partial v^{2}}\lambda(v,\phi)-\frac{3\kappa}{2\gamma(\phi)^{2}}\frac{\partial}{\partial v}\lambda(v,\phi)-\frac{\kappa^{2}}{\gamma(\phi)^{2}}\lambda(v,\phi)+\kappa\,\mathcal{F}(\phi)- T_{vv}(v,\phi) = 0. \label{7}
\end{align}
Next we consider a shock wave type stress tensor, i.e. $T_{vv}(v,\phi) = s \delta(v-v_{0})g(\phi)$. Where `$s$' is a constant or can also be a function of $\phi$.  The situation still describes a fixed temperature but not a stationary black hole configuration. 

For non-stationary background, let us first try to solve (\ref{4}) together with (\ref{5}) perturbatively in $\kappa$ to gain insight into the problem. Upto leading order in $\kappa$, we get from (\ref{4}) and (\ref{5}),
\begin{align}
\gamma^{2}(\partial_{v}^{2}S^{\phi}-\kappa\partial_{v}S^{\phi}) =- \rho\Big(\frac{1}{2}\kappa \theta (\phi)^{2}+\frac{\kappa\theta(\phi)\gamma(\phi)^{'}}{\gamma(\phi)}-\kappa\theta(\phi)^{'}+\frac{1}{2}\kappa\partial_{v}\lambda(v,\phi)-\frac{1}{2}\partial^{2}_{v}\lambda(v,\phi)\Big) S^{\phi}+\mathcal{O}(\kappa^2).
\end{align}
Integrating this twice, we get
\begin{align}
\Delta S^{\phi}(1-\kappa\,\Delta v)=\frac{\rho}{\gamma^2}\Big(\frac{1}{2}\,\Delta\lambda-\frac{1}{2}\,\kappa\,\Delta v\, \Delta\lambda-\kappa (\Delta v)^2 \Theta(\phi)\Big)S^{\phi}+\mathcal{O}(\kappa^2),
\end{align}\footnote{$\Delta v=v_f-v_i$ as defined earlier.}
where
\begin{align}
    \Theta(\phi)=\frac{\theta(\phi)^2}{2}+\frac{\theta(\phi)\gamma(\phi)^{'}}{\gamma(\phi)}-\theta(\phi)^{'}.
\end{align}
Finally, we get
\begin{align}
    \Delta S^{\phi}=\frac{\rho}{\gamma^2}\Big(\frac{1}{2}\,\Delta\lambda-\kappa (\Delta v)^2 \Theta(\phi)\Big)S^{\phi}+\mathcal{O}(\kappa^2),\label{7b}
\end{align}
Now from (\ref{7}) upto $\mathcal{O}(\kappa)$, we get integrating twice
\begin{align}
\Delta\lambda\simeq 2\gamma^2\Big[-s\,\Delta v\, g(\phi)+\kappa\,(\Delta v)^2\, (\mathcal{F}+3\,s\,g(\phi) )\Big]. \label{7a}
\end{align}
Inserting Eq. (\ref{7a}) in (\ref{7b}) describes an expression of shift in deviation vector due to passage of a gravitational wave pulse near the asymptotic horizon of a non-extreme black hole. \\

Now let us look at the connection with BMS symmetries for nonextreme case. We shall adopt the similar approach as has been shown for extreme case. The general solution for $\lambda(v,\phi)$ can be written as
\begin{align}
    \lambda(v,\phi)=c(\phi)+\tilde{g}(\phi)v+\kappa\mathcal{\tilde{F}}(\phi)v\Delta v,
\end{align}
The $v$ component of the Killing vector for the fixed temperature configuration is given by Eq.(\ref{fxdT3D}). With these considerations, the variation of $\lambda(v,\phi)$ along Killing direction gives a condition generated from the $v$ coefficient. This makes sure that the ${\cal L}_{Z}\lambda$ remains independent of $v$. The condition is given by
\begin{align}
    (2\tilde{g}+\kappa\mathcal{\tilde{F}}\Delta v)\partial_{\phi}Y+(Y\partial_{\phi}\tilde{g}+\kappa\Delta v \partial_{\phi}\mathcal{\tilde{F}})-2\kappa\theta\partial_{\phi}X+2\kappa\partial^{2}_{\phi}X-2\kappa\frac{\partial_{\phi}\gamma(\phi)}{\gamma(\phi)}\partial_{\phi}X = 0 , \label{3dnxt2}
\end{align}
where $\tilde{g}$, $\mathcal{\tilde{F}}$, $Y$, $\theta$ and $X$ are functions of $\phi$. This condition also involves the superrotation $Y$ and supertranslation $X$. Further, since $X$ does not appear in the charge of BMS symmetries \cite{Donnay2016}, then it can be considered to be zero. This simplifies the Eq. (\ref{3dnxt2}), and reduces to
\begin{align}
    (2\tilde{g}+\kappa\mathcal{\tilde{F}}\Delta v)\partial_{\phi}Y+(Y\partial_{\phi}\tilde{g}+\kappa\Delta v \partial_{\phi}\mathcal{\tilde{F}}) = 0 , \label{3dnxtn2}
\end{align}
However, except the $v$ coefficient, the rest of the terms in Eq. (\ref{sup1}) would contain only functions of $\phi$. In this way a connection between BMS parameters and displacement memory can be established.

\section{Near-Horizon Memory for Four-Dimensional Black Holes}\label{4D}
Next we consider a realistic scenario by extending our analysis to four-dimensional black holes. We shall again adopt a similar strategy as implemented in section (\ref{3D}). We consider the same configuration, i.e., two nearby timelike geodesics or test detectors are being positioned near the black hole horizon. Let us start by writing the  general form of the near-horizon four-dimensional metric \cite{PhysRevLett.116.091101,Donnay2016,PhysRevD.98.124016}
\begin{align}
ds^{2} =& g_{vv} dv^{2}+2k dv d\rho +2g_{vA} dv dx^{A}+g_{AB}dx^{A}dx^{B}, \label{m}
\end{align}
with following fall-off conditions for the horizon $\rho = 0 $:
\begin{align}
    g_{vv} =& -2\kappa\rho + \mathcal{O}(\rho^{2}) \hspace{3mm} ; \hspace{5mm} k = 1+\mathcal{O}(\rho^{2}) \nonumber\\
    g_{vA} =& \rho \theta_{A}+\mathcal{O}(\rho^{2}) \hspace{8.3mm} ; \hspace{5mm} g_{AB} = \Omega \gamma_{AB}+\rho \lambda_{AB}+\mathcal{O}(\rho^{2}) \nonumber
\end{align}
where $\theta_{A}$, $\Omega$ and $\lambda^{AB}$ are functions of ($v,x^{A}$). We shall consider $\Omega$ to be independent of $v$. $\gamma_{AB}$ represents the 2-sphere metric. In stereographic coordinates, $x^{A}=(\zeta, \bar{\zeta})$, the spherical part of the metric is $\gamma_{AB} dx^{A} dx^{B} = \frac{4}{(1+\zeta\bar{\zeta})^{2}} d\zeta d\bar{\zeta}$.
While metric components $g_{\rho A}$ and $g_{\rho\rho}$ fall as $\mathcal{O}(\rho^{2})$.  
The asymptotic Killing vectors preserving fall-off boundary conditions are given by 
\begin{align}
\chi^{v} =& f(v,x^{A}) \nonumber \\
\chi^{\rho} =& -\partial_{v}f \rho + \frac{1}{2}g^{AB}\theta_{A}\partial_{B}f\rho^{2}+\mathcal{O}(\rho^{3}) \label{killing4D}\\
\chi^{A} =& Y^{A}(x^{A})+g^{AC}\partial_{C}f \rho+\frac{1}{2}g^{AD}g^{CB}\lambda_{DB}\partial_{C}f \rho^{2}+\mathcal{O}(\rho^{3}), \nonumber 
\end{align}
where $f(v,x^{A})$ and $Y^A(x^{A})$ are arbitrary functions, and $g^{AB}$ is inverse of $g_{AB}$. $Y^{A}$ is a function of $x^{A}$ only, i.e. $Y^{\zeta} = Y^{\zeta}(\zeta)$ and $Y^{\bar{\zeta}} = Y^{\bar{\zeta}}(\bar{\zeta})$. One can find the transformation of the functions $\theta_{A}, \lambda_{AB}, g_{AB}$ and $\kappa$, and can establish the algebra between supertranslation and superrotation as an  extension of three-dimensional analysis. Further, for $\kappa$ constant together with $\partial_{v}\Omega =0$, the conserved charge at the horizon turns out to be
\begin{align}
    \mathcal{Q}(T,Y^{A}) = \frac{1}{16\pi G}\int d\zeta d\bar{\zeta}\sqrt{\gamma}\Omega  \Big(2\kappa T-Y^{A}\theta_{A}\Big). \label{chrg1d4}
\end{align}
They close under the Poisson bracket given by
\begin{align}
    \{\mathcal{Q}(T_{1},Y^{A}_{1}),\mathcal{Q}(T_{2},Y^{A}_{2})\} = \mathcal{Q}(T_{12},Y^{A}_{12}). \label{4dxtnn}
\end{align}
Similarly, one also obtains for $\kappa = 0$
\begin{align}
    \mathcal{Q}(X,Y^{A}) = \frac{1}{16\pi G}\int d\zeta d\bar{\zeta}\sqrt{\gamma}\Big(2X\Omega -Y^{A}\theta_{A}\Omega \Big)+\mathcal{Q}_{0}, \label{extchrd4d}
\end{align}
where $T$ and $X$ are set of supertranslations and $Y^{A}$ denotes superrotation. The explicit details can be found in \cite{PhysRevLett.116.091101,Donnay2016,PhysRevD.98.124016}. Next we consider the memory effect analysis for non-extreme and extreme considerations.

\subsection{Non-Extremal Case}\label{4des}

Let us examine the non-extreme case first. We again consider the fixed temperature configuration for the full metric. The deviation equation for the metric (\ref{m}) is
\begin{align}
\frac{2\Omega}{(1+\zeta\bar{\zeta})^{2}}(\partial^{2}_{v}S^{\bar{\zeta}}-\kappa\partial_{v}S^{\bar{\zeta}}) = -(R_{\zeta v\bar{\zeta} v} S^{\bar{\zeta}}+R_{\zeta v\zeta v} S^{\zeta}), \label{ddn21n}
\end{align}
where $S^{\bar{\zeta}}$ and $S^{\zeta}$ depict the $\bar{\zeta}$ and $\zeta$ components of the deviation vector and $\Omega$ is function of ($\zeta,\bar{\zeta}$). The  corresponding Riemann tensor components to the linear order in $\kappa$ and $\rho$ are given by
\begin{align}
R_{\zeta v\bar{\zeta}v} =& \frac{1}{2} \rho (\partial_{v}\partial_{\bar{\zeta}}\theta_{\zeta}+\partial_{v}\partial_{\zeta}\theta_{\bar{\zeta}}-\kappa (\partial_{\bar{\zeta}}\theta_{\zeta}-\theta_{\zeta} \theta_{\bar{\zeta}}+\partial_{\zeta}\theta_{\bar{\zeta}})+\kappa\partial_{v}\lambda_{\zeta\bar{\zeta}}-\partial^{2}_{v}\lambda_{\zeta\bar{\zeta}})+ \mathcal{O}(\rho^{2})\\
R_{\zeta v\zeta v} =& \frac{\rho}{2 (\zeta\bar{\zeta}+1) \Omega} (-2 (\zeta\bar{\zeta}+1) \partial_{\zeta}\Omega \partial_{v}\theta_{\zeta}+2\kappa ((\zeta\bar{\zeta}+1) \partial_{\zeta}\Omega-2 \bar{\zeta} \Omega) \theta_{\zeta}+\kappa (\zeta\bar{\zeta}+1) \Omega \theta_{\zeta}^2+\nonumber \\
& \Omega(-2\kappa (1 + \zeta\bar{\zeta}) \partial_{\zeta}\theta_{\zeta}+4 \bar{\zeta} \partial_{v}\theta_{\zeta}+2 (1+\zeta\bar{\zeta}) \partial_{v}\partial_{\zeta}\theta_{\zeta}+\kappa (1+\zeta\bar{\zeta})\partial_{v}\lambda_{\zeta\zeta}-(1+\zeta\bar{\zeta})\partial^{2}_{v}\lambda_{\zeta\zeta}))+\mathcal{O}(\rho^{2}). \label{rmn4dfl}
\end{align}
Therefore, the change in the deviation vector can now be written as
\begin{align}
\frac{2\Omega}{(1+\zeta\bar{\zeta})^{2}}\Big(1-\kappa\Delta v\Big)\Delta S^{\bar{\zeta}} = -\int\int(R_{\zeta v\zeta v}S^{\zeta}+R_{\zeta v\bar{\zeta}v}S^{\bar{\zeta}})dv dv . \label{gnd2}
\end{align}
Note that
\begin{align}
    \int\int R_{\zeta v\bar{\zeta}v} dv dv = \frac{\rho}{2}\Big(\Delta (\partial_{\bar{\zeta}}\theta_{\zeta})\Delta v+\Delta (\partial_{\zeta}\theta_{\bar{\zeta}})\Delta v-\kappa\mathcal{H}(\zeta,\bar{\zeta})(\Delta v)^{2}+\kappa\Delta\lambda_{\zeta\bar{\zeta}}\Delta v-\Delta\lambda_{\zeta\bar{\zeta}}\Big)+\mathcal{O}(\rho^{2}), \label{rmn4dfull}
\end{align}
with $\mathcal{H}(\zeta,\bar{\zeta})=(\partial_{\bar{\zeta}}\theta_{\zeta}-\theta_{\zeta} \theta_{\bar{\zeta}}+\partial_{\zeta}\theta_{\bar{\zeta}})$. Similarly, one can now get an expression for integral of $R_{\zeta v\zeta v}$. One can also compute the $\Delta S^{\zeta}$ component of the displacement vector adopting the similar approach. We are concentrating on $\Delta S^{\bar{\zeta}}$ component throughout the draft. In general, the Eq. (\ref{gnd2}) will contain changes `$\Delta$' of $\theta_{\zeta}$, $\lambda_{\zeta\zeta}$ and $\lambda_{\zeta\bar{\zeta}}$ together with the changes in their derivatives with respect to $v$. We can see that all metric parameters would contribute in the displacement memory. These changes in the metric parameters can be determined using Einstein field equations. The relevant component of the field equation is given by
\begin{align}
8\pi T^{M}_{vv}=& \frac{\rho}{2 \Omega^3} ((\zeta\bar{\zeta}+1)^2 \Omega^2 (-\partial_{v}\theta_{\zeta}\theta_{\bar{\zeta}}-\kappa \partial_{\bar{\zeta}}\theta_{\zeta} +\partial_{v}\partial_{\bar{\zeta}}\theta_{\zeta}+\theta_{\zeta}(2 \kappa  \theta_{\bar{\zeta}}-\partial_{v}\theta_{\bar{\zeta}})- \kappa \partial_{\zeta}\theta_{\bar{\zeta}}+\partial_{v}\partial_{\zeta}\theta_{\bar{\zeta}}+\kappa \partial_{v}\lambda_{\zeta\bar{\zeta}}-\partial^{2}_{v}\lambda_{\zeta\bar{\zeta}})\nonumber \\
& -\kappa (\Omega^2 (3 (\zeta\bar{\zeta}+1)^2 \theta_{\zeta} \theta_{\bar{\zeta}}-2 ((\zeta\bar{\zeta}+1)^2 \partial_{\bar{\zeta}}\theta_{\zeta}+(\zeta\bar{\zeta}+1)^2 \partial_{\zeta}\theta_{\bar{\zeta}}- 2 ((\zeta\bar{\zeta}+1)^2 \partial_{v}\lambda_{\zeta\bar{\zeta}}-1))+\nonumber \\
& 4 \kappa (\zeta\bar{\zeta}+1)^2 \lambda_{\zeta\bar{\zeta}})-2 (\zeta\bar{\zeta}+1)^2 \partial_{\bar{\zeta}}\Omega \partial_{\zeta}\Omega+2 (\zeta\bar{\zeta}+1)^2 \Omega \partial_{\zeta}\partial_{\bar{\zeta}}\Omega))+\mathcal{O}(\rho^{2}). \label{efe4dfl}
\end{align}
Considering again the shockwave profile of the stress-tensor as $T_{vv}=\lim_{\rho\to 0}\frac{8\pi T^{M}_{vv}}{\rho}$, we could determine the required changes for the Eq. (\ref{gnd2}). Now, using the $vA$-component of the Einstein field equations, we get the following conditions to the $\mathcal{O}(\rho^{0})$  
\begin{align}
    \partial_{v}\theta_{A} = 0 \hspace{2mm} \Rightarrow \hspace{2mm} \theta_{A} = \mathcal{C}(x^{A}). \label{thtacondn}
\end{align}
Further, for computational simplification, we also consider $\Omega$ to be unity. Under these considerations, the calculation reduces significantly. The Riemann tensor components are given by
\begin{align}
\int\int R_{\zeta v\zeta v} dv dv = \frac{\rho}{2}\Big(\Delta\lambda_{\zeta\zeta} (\kappa\Delta v-1)-\kappa \mathcal{G}(\zeta,\bar{\zeta})(\Delta v)^{2}\Big), \label{4dr1}
\end{align}
where $ \mathcal{G}(\zeta,\bar{\zeta})=2\partial_{\zeta}\theta_{\zeta}+\theta_{\zeta}^{2}+\frac{4\bar{\zeta}\theta_{\zeta}}{(1+\zeta\bar{\zeta})} $, and
\begin{align}
\int\int R_{\zeta v\bar{\zeta}v} dv dv = \frac{\rho}{2}\Big(\Delta\lambda_{\zeta\bar{\zeta}}(\kappa\Delta v-1)-\kappa\mathcal{H}(\zeta,\bar{\zeta})(\Delta v)^{2}\Big). \label{4dr2}
\end{align}
This enables us to rewrite the GDE solution (\ref{gnd2}) in a compact form, 
\begin{align}
    \Delta S^{\bar{\zeta}} = \frac{\rho (1+\zeta\bar{\zeta})^{2}}{4}\Big((\Delta\lambda_{\zeta\bar{\zeta}}S^{\bar{\zeta}}+\Delta\lambda_{\zeta\zeta}S^{\zeta})-\kappa (\Delta v)^{2}(\mathcal{H}S^{\bar{\zeta}}+\mathcal{G}S^{\zeta})\Big)+\mathcal{O}(\rho^{2}). \label{ngd3n}
\end{align}
Thus we now only need to find the changes in $\lambda_{AB}$ in order to solve the Eq. (\ref{ngd3n}). This requires to solve the EFE as we did for three dimensional case. The $vv$-component EFE is
\begin{align}
8\pi T^{M}_{vv} = \frac{\rho}{2}  (\zeta\bar{\zeta}+1)^2 (\kappa (-3 \partial_{v}\lambda_{\zeta\bar{\zeta}}+\partial_{\bar{\zeta}}\theta_{\zeta}-\theta_{\zeta} \theta_{\bar{\zeta}}+\partial_{\zeta}\theta_{\bar{\zeta}})-\partial^{2}_{v}\lambda_{\zeta\bar{\zeta}}+\frac{4\kappa }{(\zeta\bar{\zeta}+1)^2})+\mathcal{O}(\rho^{2}). \label{nefe4d}
\end{align}
Now again defining $T_{vv}=\lim_{\rho\to 0}\frac{8\pi T^{M}_{vv}}{\rho}=aQ(\zeta,\bar{\zeta})\delta(v-v_{0})$, with some constant $a$, we get
\begin{align}
\Delta\lambda_{\zeta\bar{\zeta}} = \Big(\kappa\tilde{G}\Delta v-\frac{2aQ}{(1+\zeta\bar{\zeta})^{2}}\Big)\Delta v,
\end{align}
where $\tilde{G}(\zeta,\bar{\zeta})=\mathcal{H}(\zeta,\bar{\zeta})+\frac{4}{(1+\zeta\bar{\zeta})^{2}}+\frac{6a\Delta v}{(1+\zeta\bar{\zeta})^{2}}Q(\zeta,\bar{\zeta})$. Further, for determining $\Delta\lambda_{\zeta\zeta}$, we use $\zeta\zeta$-component of the field equation. We get an $\mathcal{O}(\rho^{0})$ term together with a linear order correction in $\rho$. The $\mathcal{O}(\rho^{0})$ term gives us
\begin{align}
\partial_{v}\lambda_{\zeta\zeta}+\kappa\lambda_{\zeta\zeta}=\frac{2\bar{\zeta}\theta_{\zeta}}{(1+\zeta\bar{\zeta})}-\frac{1}{2}\theta_{\zeta}^{2}. \label{nefe4dr}
\end{align}
This generates the change in $\lambda_{\zeta\zeta}$ with $\tilde{A}= \frac{2\bar{\zeta}\theta_{\zeta}}{(1+\zeta\bar{\zeta})}-\frac{1}{2}\theta_{\zeta}^{2}$.
\begin{align}
    \lambda_{\zeta\zeta} = \frac{1}{\kappa}\tilde{A}+C_{1}(\zeta,\bar{\zeta})(1-v\kappa) \Rightarrow \Delta\lambda_{\zeta\zeta} = \kappa \tilde{B}(\zeta,\bar{\zeta})\Delta v ,
\end{align}
where $C_{1}(\zeta,\bar{\zeta})$ appears as a integration constant. As $\Delta\lambda_{\zeta\zeta}$ and $\Delta\lambda_{\zeta\bar{\zeta}}$ are contained in Eq. (\ref{ngd3n}), we can replace these changes to obtain an explicit form of the memory. 
\begin{align}
    \Delta S^{\bar{\zeta}} = \frac{\rho (1+\zeta\bar{\zeta})^{2}}{4}\Big(\Big((\kappa\tilde{G}\Delta v-\frac{2aQ}{(1+\zeta\bar{\zeta})^{2}}) S^{\bar{\zeta}}+\kappa \tilde{B}S^{\zeta}\Big)\Delta v-\kappa (\Delta v)^{2}(\mathcal{H}S^{\bar{\zeta}}+\mathcal{G}S^{\zeta})\Big)+\mathcal{O}(\rho^{2}). \label{ngd3nn}
\end{align}
We observe that there is a direct relation between $\lambda_{AB}$ and $\theta_{A}$  from Eq. (\ref{nefe4d}) or Eq. (\ref{nefe4dr}). This also suggests that the displacement memory is ultimately restored in the changes of metric parameters $\theta_{A}$. This completes our analysis of achieving the displacement memory for near the horizon of non-extremal black holes. \\


\textbf{Relation with BMS Symmetry}\\ \\
Let us now establish the connection between displacement memory and BMS symmetries in the vicinity of black hole horizon generated by shockwave profile.  
We have considered here the transformations in $\lambda_{\zeta\bar{\zeta}}$, $\lambda_{\zeta\zeta}$ and $\lambda_{\bar{\zeta}\bar{\zeta}}$. Hence, the general solution for $\lambda_{\zeta\bar{\zeta}}$, $\lambda_{\zeta\zeta}$ and $\lambda_{\bar{\zeta}\bar{\zeta}}$ can be written as 
\begin{align}
    \lambda_{\zeta\bar{\zeta}} = A(\zeta,\bar{\zeta})+B(\zeta,\bar{\zeta})v+\kappa\tilde{\mathcal{K}}(\zeta,\bar{\zeta})v\Delta v \hspace{1mm} ; \hspace{1mm} \lambda_{\zeta\zeta} = \tilde{A}(\zeta,\bar{\zeta})+v\tilde{B}(\zeta,\bar{\zeta}) \kappa \hspace{1mm};\hspace{1mm} \lambda_{\bar{\zeta}\bar{\zeta}}=\tilde{W}(\zeta,\bar{\zeta})+v\kappa\tilde{Z}(\zeta,\bar{\zeta}). \label{nd4}
\end{align}
Therefore, as before the changes in various components of $\lambda_{AB}$ are given by
\begin{align}
\Delta \lambda_{\zeta\bar{\zeta}}= B(\zeta,\bar{\zeta})\Delta v+\kappa \tilde{\mathcal{K}}(\zeta,\bar{\zeta})(\Delta v)^{2} \hspace{2mm} ; \hspace{2mm} \Delta\lambda_{\zeta\zeta}=\kappa\tilde{B}(\zeta,\bar{\zeta})\Delta v \hspace{2mm} ; \hspace{2mm} \Delta\lambda_{\bar{\zeta}\bar{\zeta}}=\kappa\tilde{Z}(\zeta,\bar{\zeta})\Delta v . \label{dlmbda2} 
\end{align}
Recall $\kappa$ is small, and for a small $\Delta v$ we can disregard the term $\kappa (\Delta v)^2$. In this case, we are  considering only the part of $\Delta S^{\bar{\zeta}}$ in (\ref{ngd3nn}) that is induced by the changes in $\lambda_{AB}$. \par

Now consider the variation of $\lambda_{AB}(v,\zeta,\bar{\zeta})$ along killing direction $\chi$ given by \cite{PhysRevD.98.124016} 
\begin{align}
{\cal L}_{\chi}\lambda_{AB} = f \partial_{v}\lambda_{AB}-\lambda_{AB}\partial_{v}f+\mathcal{L}_{Y}\lambda_{AB}+\theta_{A}\partial_{B}f+\theta_{B}\partial_{A}f-2\nabla_{A}\nabla_{B}f , \label{chi2}
\end{align}
where $\chi$ is the Killing vector. For non-extreme case, the $v$ component of the Killing vector for the fixed temperature configuration generates two sets of supertranslations $T(\phi)$ and $X(\phi)$,
\begin{align}
    f(v,\zeta,\bar{\zeta}) = T(\zeta,\bar{\zeta})+(1-\kappa v)X(\zeta,\bar{\zeta})+\mathcal{O}(\kappa^{2}). \label{k4d}
\end{align}
Now in order to make the variation of different components of $\lambda_{AB}$ along the killing direction independent of $v$, we again set the $v$ coefficients to be zero. It is apparent from Eq. (\ref{nd4}) and Eq. (\ref{k4d}) the terms dependent on $v$ will only be linear in $v$. Considering $\mathcal{L}_{\chi}\lambda_{\zeta\bar{\zeta}}$, we get the following condition
\begin{align}
    \partial_{C}\Big((B+\tilde{\mathcal{K}}\kappa\Delta v)Y^{C}\Big)-(\theta_{\zeta}\partial_{\bar{\zeta}}X+\theta_{\bar{\zeta}}\partial_{\zeta}X)\kappa+2\kappa\nabla_{\zeta}\nabla_{\bar{\zeta}}X = 0. \label{bms4d1}
\end{align}
Similarly, considering $\mathcal{L}_{\chi}\lambda_{\zeta\zeta}$, we obtain
\begin{align}
    \kappa\Big( Y^{C}\partial_{C}\tilde{B}+2\tilde{B}\partial_{\zeta}Y^{\zeta}-2\theta_{\zeta}\partial_{\zeta}X+2\nabla_{\zeta}\nabla_{\zeta}X\Big) = 0, \label{bms4d2}
\end{align}
and $\mathcal{L}_{\chi}\lambda_{\bar{\zeta}\bar{\zeta}}$ gives
\begin{align}
    \kappa\Big(Y^{C}\partial_{C}\tilde{Z}+2\partial_{\bar{\zeta}}Y^{\bar{\zeta}}\tilde{Z}-2\theta_{\bar{\zeta}}\partial_{\bar{\zeta}}X-\nabla_{\bar{\zeta}}\nabla_{\bar{\zeta}}X\Big)=0. \label{bms4dnxt3}
\end{align}
In principle Eq. (\ref{bms4d1}), Eq. (\ref{bms4d2}) and Eq. (\ref{bms4dnxt3}) can be solved for $Y^{\zeta}$, $Y^{\bar{\zeta}}$ and $X$. These equations appear as conditions which make sure that the variation of $\lambda_{AB}$ will be independent of $v$. Further, if one freezes the supertranslation $X$\footnote{Since $X$ is absent in the leading order charge Eq.(\ref{chrg1d4}), this is a legitimate assumption. }, this reduces the above conditions in the following form
\begin{align}
    \partial_{C}\Big((B+\tilde{\mathcal{K}}\kappa\Delta v)Y^{C}\Big) = 0 \hspace{2mm};\hspace{2mm}
    \kappa( Y^{C}\partial_{C}\tilde{B}+2\tilde{B}\partial_{\zeta}Y^{\zeta}) = 0 \hspace{2mm};\hspace{2mm}
    \kappa\Big(Y^{C}\partial_{C}\tilde{Z}+2\partial_{\bar{\zeta}}Y^{\bar{\zeta}}\tilde{Z}\Big)=0 .
\end{align}
Using these set of equations one can obtain the solution for $Y^{\zeta}$
\begin{align}
    Y^{\zeta}(\zeta) = \tilde{a}e^{-\int\frac{\tilde{w}}{\tilde{p}}d\zeta} , \label{intnxt4d}
\end{align}
where $\tilde{w}$ and $\tilde{p}$ are functions of ($\zeta,\bar{\zeta}$) and $\tilde{a}$ appears as an integration constant. $\tilde{a}$ will be a function of $\bar{\zeta}$ only. Similarly, one can also find the solution for $Y^{\bar{\zeta}}(\bar{\zeta})$. 
Therefore, we can explicitly find some $Y^{A}$ that would induce the desired shift in the deviation vector components.
\subsection{Extremal Case}\label{nextml2}
Next we analyze the extremal case in which the $g_{vv}$ component of the metric becomes $\mathcal{N}(\zeta,\bar{\zeta})\rho^{2}+\mathcal{O}(\rho^{3})$. We obtain the GDE as
\begin{align}
\frac{2}{(1+\zeta\bar{\zeta})^{2}}\partial^{2}_{v}S^{\bar{\zeta}}_{E} = -(R_{\zeta v\bar{\zeta} v} S^{\bar{\zeta}}_{E}+R_{\zeta v\zeta v} S^{\zeta}_{E}), \label{ddn21}
\end{align}
where $S^{\bar{\zeta}}_{E}$ and $S^{\zeta}_{E}$ depict the $\bar{\zeta}$ and $\zeta$ components of the deviation vector respectively for extremal case. The relevant components of the Riemann tensor are
\begin{align}
R_{\zeta v\bar{\zeta} v} = -\frac{\rho}{2}  \partial^{2}_{v}\lambda_{\zeta\bar{\zeta}} + \mathcal{O}(\rho^{2}) \hspace{2mm} ; \hspace{2mm} R_{\zeta v\zeta v} = -\frac{\rho}{2} \partial^{2}_{v}\lambda_{\zeta\zeta} + \mathcal{O}(\rho^{2}). \label{R1}
\end{align}
Therefore, the change in the deviation vector can now be written as
\begin{align}
\Delta S^{\bar{\zeta}}_{E} =\frac{\rho}{4}(1+\zeta\bar{\zeta})^{2}(\Delta\lambda_{\zeta\bar{\zeta}} S^{\bar{\zeta}}_{E}+\Delta\lambda_{\zeta\zeta} S^{\zeta}_{E})+\mathcal{O}(\rho^{2}). \label{ngd2}
\end{align}
We again consider the shockwave profile of stress tensor as mentioned in the non-extreme case.  The relevant components of the EFE can be obtained by setting $\kappa=0$ in Eq. (\ref{ngd3n}) and Eq. (\ref{nefe4dr}) to give,
\begin{align}
    \Delta\lambda_{\zeta\bar{\zeta}} = \mathcal{P}(\zeta,\bar{\zeta})\Delta v \hspace{2mm} ; \hspace{2mm} \Delta\lambda_{\zeta\zeta} = \tilde{C}(\zeta,\bar{\zeta})\Delta v . \label{ext4ddlbda}
\end{align}
Eq. (\ref{ngd2}) and Eq. (\ref{ext4ddlbda}) indicate that we get a finite jump in the deviation vector depicting displacement memory. 

Here also we may seek for BMS symmetry that induces the shift in $S^{A}_{E}$. This is again in principle be determined from (\ref{chi2}) by choosing $\lambda_{AB}$ linear in $v$. The entire change in $\lambda_{AB}$ then would be determined by functions that are dependent on $\zeta,\bar \zeta$ only. Further, for extreme case, the $\chi^{v}$ component contains two set of supertranslations, i.e. $f(v,\zeta,\bar{\zeta})=T(\zeta,\bar{\zeta})+vX(\zeta,\bar{\zeta})$. Let us take the general solution of $\lambda_{AB}$ for extremal consideration of the following form 
\begin{align}
    \lambda_{\zeta\bar{\zeta}} = \mathcal{A}(\zeta,\bar{\zeta})+v\mathcal{P}(\zeta,\bar{\zeta}) \hspace{2mm} ; \hspace{2mm} \lambda_{\zeta\zeta} = \mathcal{B}(\zeta,\bar{\zeta})+v\tilde{C}(\zeta,\bar{\zeta}). \label{lbdsol4d}
\end{align}
The $\bar{\zeta}\bar{\zeta}$-component of the EFE will generate a $\lambda_{\bar{\zeta}\bar{\zeta}}$, similar to $\lambda_{\zeta\zeta}$. Again, we set $v$ coefficient to be zero in order to make the variation of $\lambda_{AB}$ along the killing direction to be independent of $v$. For $\lambda_{\zeta\bar{\zeta}}$ we obtain
\begin{align}
    \partial_{C}(Y^{C}\mathcal{P})-(\theta_{\zeta}\partial_{\bar{\zeta}}X+\theta_{\bar{\zeta}}\partial_{\zeta}X)-2\nabla_{\zeta}\nabla_{\bar{\zeta}} X = 0. \label{cndn2n4d}
\end{align}
and for $\lambda_{\zeta\zeta}$ variation
\begin{align}
    Y^{C}\partial_{C}\tilde{C}+2\tilde{C}\partial_{\zeta}Y^{\zeta}-2\theta_{\zeta}\partial_{\zeta}X-2\nabla_{\zeta}\nabla_{\zeta}X = 0. \label{cndn4d2n}
\end{align}
Similarly, variation in $\lambda_{\bar{\zeta}\bar{\zeta}}$ also generates a equation similar to Eq. (\ref{cndn4d2n}). These conditions should serve the purpose to obtain explicit solutions for $X$, $Y^{\zeta}$ and $Y^{\bar{\zeta}}$ that can generate variations in $\lambda_{AB}$ independent of $v$. However, obtaining an explicit solution for $X$ or $Y$ seems to be difficult in this case. Some insights can be gained considering a reduced form of the metric which we study in the next section.


\subsection{Memory for a less generic metric}\label{rdcdm}

In this section, we consider a reduced form of the metric (\ref{m}) where we set $g_{vA}=0$. This can be regraded as an asymptotic form of a metric near the horizon of a spherically symmetric black hole which is deformed in the spatial sector. The metric now reads
\begin{align}
  ds^{2} =& g_{vv} dv^{2}+2k dv d\rho+g_{AB}dx^{A}dx^{B}. \label{m1}
\end{align}

\subsubsection{Non-Extremal Case}\label{rdcdnxtm}

We consider the non-extreme case first for the above form of the metric. We shall be adopting the similar approach for computing the displacement memory as we have shown in section (\ref{4des}). We consider again the fixed temperature configuration, i.e. $\kappa$ to be constant. Let us write down the GDE for the metric (\ref{m1})
\begin{align}
\frac{2}{(1+\zeta\bar{\zeta})^{2}}(\partial^{2}_{v}S^{\bar{\zeta}}-\kappa\partial_{v}S^{\bar{\zeta}}) = -(R_{\zeta v\bar{\zeta} v} S^{\bar{\zeta}}+R_{\zeta v\zeta v} S^{\zeta}), \label{ddn21nn}
\end{align}
Therefore, the change in the deviation vector can now be written as
\begin{align}
\Delta S^{\bar{\zeta}} =\frac{\rho}{4}(1+\zeta\bar{\zeta})^{2}(\Delta\lambda_{\zeta\bar{\zeta}} S^{\bar{\zeta}}+\Delta\lambda_{\zeta\zeta} S^{\zeta})+\mathcal{O}(\rho^{2}). \label{ngd2nn}
\end{align}
The $vv$-component of the EFE is given by
\begin{align}
- \frac{1}{2}\left(\zeta\bar{\zeta} + 1\right)^{2} \Big(\left( 3\kappa \partial_{v} \lambda_{\zeta\bar{\zeta}} + \partial^{2}_{v} \lambda_{\zeta\bar{\zeta}} \right) + 2 \kappa\Big)\rho  - T^{M}_{vv}{\left(v,\zeta, \bar{\zeta} \right)} = 0. \label{En1n}
\end{align}
With a shockwave profile $T_{vv}=\lim_{\rho\to 0}\frac{8\pi T^{M}_{vv}}{\rho}=aQ(\zeta,\bar{\zeta})\delta(v-v_{0})$, for some constant $a$, we get
\begin{align}
\Delta \lambda_{\zeta\bar \zeta} = -\frac{2\, a}{(1+\zeta\bar \zeta)^2}\Delta v\,(1- 3\,\kappa\,\Delta v)\, Q(\zeta,\bar \zeta)+\frac{4\,\kappa\,(\Delta v)^2}{(1+\zeta\bar \zeta)^2}. \label{RR4nn}
\end{align}
Further, the $\zeta\zeta$-component of the EFE one gets
\begin{align}
\Delta\lambda_{\zeta\zeta}=\kappa\mathcal{W}(\zeta,\bar{\zeta})\Delta v .\label{lbda2rdcn}    
\end{align}
Hence, from Eq. (\ref{ngd2nn}), Eq. (\ref{RR4nn}) and Eq. (\ref{lbda2rdcn}), we find that the memory can be associated with $\Delta\lambda_{\zeta\bar{\zeta}}$ and $\Delta\lambda_{\zeta\zeta}$ to the linear order in $\rho$. \\

\textbf{Relation with BMS Symmetry}\\ \\
To relate it with BMS symmetry, we have the general solutions of $\lambda_{AB}$ of the same form as it appears for non-extreme case in section (\ref{4des}). 
\begin{align}
    \lambda_{\zeta\bar{\zeta}} = A(\zeta,\bar{\zeta})+B(\zeta,\bar{\zeta})v+\kappa\mathcal{C}(\zeta,\bar{\zeta})v\Delta v \hspace{2mm} ; \hspace{2mm} \lambda_{\zeta\zeta} = \tilde{A}(\zeta,\bar{\zeta})+v\tilde{B}(\zeta,\bar{\zeta}) \kappa \label{lbdardcd4d}
\end{align}
This generates the same conditions as appeared in section (\ref{4des}). We obtain the similar form of the solution as mentioned in Eq. (\ref{intnxt4d}).

\subsubsection{Extremal Case}\label{xnt4nd}
Let us now consider the extremal case. As we have seen the extremal analysis for the full metric in (\ref{nextml2}), the $g_{vv}$ component would change by a function of ($\zeta,\bar{\zeta}$) to the $\rho^{2}$ order. Under this consideration, the deviation equation for the metric (\ref{m1}) reads
\begin{align}
\frac{2}{(1+\zeta\bar{\zeta})^{2}}\partial^{2}_{v}S^{\bar{\zeta}}_{E} = -(R_{\zeta v\bar{\zeta} v} S^{\bar{\zeta}}_{E}+R_{\zeta v\zeta v} S^{\zeta}_{E}), \label{dd21}
\end{align}
The change in the deviation vector can now be written as
\begin{align}
\Delta S^{\bar{\zeta}}_{E} =\frac{\rho}{4}(1+\zeta\bar{\zeta})^{2}(\Delta\lambda_{\zeta\bar{\zeta}} S^{\bar{\zeta}}_{E}+\Delta\lambda_{\zeta\zeta} S^{\zeta}_{E})+\mathcal{O}(\rho^{2}). \label{gd2}
\end{align}
From the $vv$-component of EFE we get 
\begin{align}
-\frac{\rho}{2}(1+\zeta\bar{\zeta})^{2}\partial^{2}_{v}\lambda_{\zeta\bar{\zeta}} = 8\pi T^{M}_{vv}. \label{ffd2}
\end{align}
Using the same definition of stress-energy tensor as mentioned in non-extreme case, we get 
\begin{align}
\Delta \lambda_{\zeta\bar{\zeta}} = H(\zeta,\bar{\zeta})\Delta v . \label{n4dnew}
\end{align}
From $\zeta\zeta$-component of the EFE to the $\mathcal{O}(\rho^{0})$, $\lambda_{\zeta\zeta}$ turns out to be a function of $(\zeta,\bar{\zeta})$ only. Hence, there will be no change induced by $\lambda_{\zeta\zeta}$, and Eq. (\ref{gd2}) becomes
\begin{align}
    \Delta S^{\bar{\zeta}}_{E} =\frac{\rho}{4}(1+\zeta\bar{\zeta})^{2}H(\zeta,\bar{\zeta}) \Delta v S^{\bar{\zeta}}_{E}+\mathcal{O}(\rho^{2}). \label{mxt4d}
\end{align}
Let us now look at the relation with BMS symmetry for extremal consideration. It is to note that $\mathcal{O}(\rho^{0})$ term of $\bar{\zeta}\bar{\zeta}$-component of the EFE also turns out to be function of $(\zeta,\bar{\zeta})$ only. Hence, the change in diagonal components of $\lambda_{AB}$ with respect to $v$ does not contribute in the memory. However, general solutions of $\lambda_{\zeta\zeta}$ and $\lambda_{\bar{\zeta}\bar{\zeta}}$ can still contribute while relating to BMS symmetry. The general solution for the off-diagonal component of $\lambda_{AB}$ can be written as $\lambda_{\zeta\bar{\zeta}}(v,\zeta,\bar{\zeta}) = \tilde{P}(\zeta,\bar{\zeta})+vH(\zeta,\bar{\zeta})$. While $\lambda_{\zeta\zeta}$ and $\lambda_{\bar{\zeta}\bar{\zeta}}$ are of the form $\lambda_{\zeta\zeta}=\tilde{E}(\zeta,\bar{\zeta})$ and $\lambda_{\bar{\zeta}\bar{\zeta}}=\tilde{W}(\zeta,\bar{\zeta})$. To have the desired change as depicted in Eq. (\ref{n4dnew}), we obtain the following condition for $\lambda_{\zeta\bar{\zeta}}$
\begin{align}
    \partial_{C}(Y^{C}H)-2\nabla_{\zeta}\nabla_{\bar{\zeta}}X = 0. \label{diff4d2}
\end{align}
This can further be written as
\begin{align}
    \partial_{C}(Y^{C}H)-2\Big(\partial_{\zeta}\partial_{\bar{\zeta}}X-(\Gamma^{\zeta}{}_{\zeta\bar{\zeta}}\partial_{\zeta}X+\Gamma^{\bar{\zeta}}{}_{\zeta\bar{\zeta}}\partial_{\bar{\zeta}}X)\Big) = 0, \label{diff4d22}
\end{align}
where
\begin{align}
    \Gamma^{\zeta}{}_{\zeta\bar{\zeta}}=\frac{\rho}{4} (1+\zeta\bar{\zeta})^{2}\partial_{\zeta}\lambda_{\bar{\zeta}\bar{\zeta}} +\mathcal{O}(\rho^{2}) \hspace{2mm} ; \hspace{2mm} \Gamma^{\bar{\zeta}}{}_{\zeta\bar{\zeta}} = \frac{\rho}{4} (1+\zeta\bar{\zeta})^{2}\partial_{\bar{\zeta}}\lambda_{\zeta\zeta}+\mathcal{O}(\rho^{2}). \label{chrst}
\end{align}
To the order $\mathcal{O}(\rho^{0})$, the Eq. (\ref{diff4d22}) becomes
\begin{align}
    \partial_{C}(Y^{C}H)-2\partial_{\zeta}\partial_{\bar{\zeta}}X = 0. \label{rd4d}
\end{align}
This looks to be a simpler condition to be satisfied by $X$ and $Y^{A}$. Still, getting explicit expressions is not an easy task. However, if we freeze the superrotation of the Eq. (\ref{rd4d}) reduces to,  $\partial_{\bar{\zeta}}\partial_{\zeta}X(\zeta,\bar{\zeta})=0$ \footnote{This is legitimate as the 4D charge expression contains supertranslation and superrotation separately Eq. (\ref{extchrd4d})}. For which, we can obtain an exact solution, i.e., $X(\zeta,\bar{\zeta})=X_{1}(\zeta)+X_{2.}(\bar{\zeta})$. Hence, there is a  supertranslation $X(\zeta,\bar{\zeta})$ that can induce the displacement memory effect depicted in Eq. (\ref{mxt4d}). We also notice here, the form of the equation (\ref{ffd2}) is quite similar that was obtained in the far region in \cite{Strominger2017}. The extremal case for this reduced metric mimics the memory or shift in the far region  \cite{PhysRevD.98.124016,Fernandes:2020jto}.

 Finally, we would like to mention here that the metric considered here generates gravitational waves which can be captured in the Newman-Penrose (NP) scalar $\psi_4.$ This is displayed in the appendix (\ref{exactan}). Although $\lambda_{AB}$ are not explicitly present in the charge  Eq. (\ref{extchrd4d}) but $\psi_4$ contains $\lambda_{AB}$ and their derivatives, ensuring the physicality and relevance of $\lambda_{AB}$. 

\section{Discussion}\label{dscsn}
The primary objective of this report is to study how GWs affect inertial test detectors near the horizon of black hole spacetimes. This analysis could serve as a model for estimating the deviation between the test masses of a GW detector. We also provide an explicit connection between BMS symmetries and memory effect near the horizon of black holes.\par
There are crucial differences between memory effect and asymptotic symmetries obtained in the future null infinities and the same obtained near the horizon of a black hole. As the algebra of asymptotic symmetries contains two copies of Virasoro algebra, we have two sets of supertranslation and a set of superrotation generators near the horizon of a 4d black hole. The extra set of supertranslation generator is something that we do not recover at the null infinities. Another crucial difference seems to be in the structure of the geodesic deviation equation (GDE). The near-horizon GDE contains a linear derivative term of the deviation vector with respect to time apart from a double-time derivative, whereas the far region GDE contains only a double-time derivative of deviation vector to the leading order. The function $\lambda(\lambda _{AB}$ in four dimensions) mimics the derivative of News tensor, $C_{zz}$, in the far region.  The Riemann tensor in GDE also contains a double derivative and a single derivative of $\lambda$, which is absent in the GDE of far region (\ref{asymdev}). These differences make the analyses of near-horizon memory significantly non-trivial than the far region case. We also see the memory in the near region of black holes is due to the dual effect of supertranslation and superrotation generators. For the sake of getting an explicit expression of BMS generators that would generate the displacement memory, we have considered one type of generator (either superrotations or supertranslations) at a time for extremal cases. Although it was not easy to obtain an explicit solution but   considering a slightly less generic metric, we could recover closed form solutions for BMS parameters. For this extremal case, the GDE governing the displacement memory becomes almost identical to the structure of the same in the far region.  The full non-extremal case seems to be analytically difficult to analyze. We have provided an analytic expression for the memory effect in small $\kappa$ approximation. We wish to report an analysis without such approximation in future.     \par 

In the near future, the observational aspects of GW will serve a great purpose in measuring the memory effect, and this could allow us to examine BMS symmetries in more depth as a direct evidence \cite{Khera:2020mcz}. The detection of near-horizon memory in the far region may be possible if the test detectors or particles near the black hole region interact with some matter or radiation that can be detected in the far region and encode the near-horizon physics including memory. It will be interesting to investigate the existence of such a phenomenon.
\section*{Acknowledgements} 
 S. B. is supported by SERB-DST through the Early Career Research award grant no. ECR/2017/002124 and by IIIT Allahabad through seed grant (File number 10497). A.B. is supported by Research Initiation Grant (RIG/0300) provided by IIT-Gandhinagar and Start-Up Research Grant (SRG/2020/001380) by Department of Science \& Technology Science and Engineering Research Board (India).

\appendix

\section{Newman-Penrose Tetrad and Weyl Scalar} \label{exactan}

Here we provide the explicit expressions for Newman-Penrose (NP) tetrad together with Weyl scalar. The NP formalism  is a tetrad formalism based on a set of four null vectors. There are two real vectors denoted as $l^{a}$ and $n^{a}$, and two complex conjugate vectors $m^{a}$ and $\bar{m}^{a}$. For computational convenience, we are considering the reduced form of the metric (\ref{m1}), which is independent of $\theta_{A}(x^{A})$. We provide $\Psi_{4}$ component of the Weyl scalar for non-extreme and extreme cases separately by constructing the null tetrad. Let us mention the the normalizaiton conditions for NP tetrad are given by
\begin{align}
-l^{a}n_{a} =& m^{a}\bar{m}_{a}=1 \\
l^{a}m_{a} =& l^{a}\bar{m}_{a}=n^{a}m_{a}=n^{a}\bar{m}_{a} = 0\\
l^{a}l_{a} =& n^{a}n_{a}=m^{a}m_{a}=\bar{m}_{a}\bar{m}_{a} = 0
\end{align}
Here, $n^{\alpha}$ is nothing but $n^{\alpha}_{[NP]}$.

\subsection{Weyl scalar for the less generic metric: Non-Extremal Case}

We consider the metric (\ref{m1}) for non-extreme case. The null tetrad for this are given by
\begin{align}
l^{a} = \frac{1}{\sqrt{2}}(0,1,0,0) \hspace{3mm} ; \hspace{3mm} n^{a} = \frac{2}{\sqrt{2}}(-1,-2\kappa\rho,0,0)
\end{align}
and other two complex null tetrad
\begin{align}
m^{a} = \frac{1}{\sqrt{2}}(0,0,m_{1},m_{2}) \hspace{2mm} ; \hspace{2mm}
\bar{m}^{a} = \frac{1}{\sqrt{2}}(0,0,m_{1},\bar{m}_{2})
\end{align}
Here $m_{2}$ and $\bar{m_{2}}$ will be complex conjugate of each other. And
\begin{align}
m_{1}=-\frac{\sqrt{\rho } (\zeta\bar{\zeta}+1)^2 \sqrt{\lambda _{\bar{\zeta}\bar{\zeta}}}}{\sqrt{\rho ^2 (\zeta\bar{\zeta}+1)^4 \lambda _{\bar{\zeta}\bar{\zeta}} \lambda _{\zeta\zeta}-\left(\rho  (\zeta\bar{\zeta}+1)^2 \lambda _{\zeta\bar{\zeta}}+2\right){}^2}} 
\end{align}
\begin{align}
m_{2} =& i\frac{-\sqrt{\left(\rho ^2 (\zeta\bar{\zeta}+1)^4 \lambda _{\bar{\zeta}\bar{\zeta}} \lambda _{\zeta\zeta}-\left(\rho  (\zeta\bar{\zeta}+1)^2 \lambda _{\zeta\bar{\zeta}}+2\right){}^2\right){}^3}}{\sqrt{\rho } \sqrt{\lambda _{\bar{\zeta}\bar{\zeta}}} \left(\rho ^2 (\zeta\bar{\zeta}+1)^4 \lambda _{\bar{\zeta}\bar{\zeta}} \lambda _{\zeta\zeta}-\left(\rho  (\zeta\bar{\zeta}+1)^2 \lambda _{\zeta\bar{\zeta}}+2\right){}^2\right){}^{3/2}}+\nonumber \\
& \frac{\rho  (\zeta\bar{\zeta}+1)^2 \left(\rho  (\zeta\bar{\zeta}+1)^2 \left(\lambda _{\bar{\zeta}\bar{\zeta}} \lambda _{\zeta\zeta} \left(\rho  (\zeta\bar{\zeta}+1)^2 \lambda _{\zeta\bar{\zeta}}+2\right)+\lambda _{\zeta\bar{\zeta}}{}^2 \left(-\rho  (\zeta\bar{\zeta}+1)^2 \lambda _{\zeta\bar{\zeta}}-6\right)\right)-12 \lambda _{\zeta\bar{\zeta}}\right)-8}{\sqrt{\rho } \sqrt{\lambda _{\bar{\zeta}\bar{\zeta}}} \left(\rho ^2 (\zeta\bar{\zeta}+1)^4 \lambda _{\bar{\zeta}\bar{\zeta}} \lambda _{\zeta\zeta}-\left(\rho  (\zeta\bar{\zeta}+1)^2 \lambda _{\zeta\bar{\zeta}}+2\right){}^2\right){}^{3/2}}
 \end{align}
 Using the non-vanishing Weyl tensor components, we compute the desired Newman-Penrose scalar. The Weyl scalar $\Psi_{4}$ is given by
\begin{align}
\Psi_{4} = -C_{\alpha\beta\mu\nu}n^{\alpha}\bar{m}^{\beta}n^{\mu}\bar{m}^{\nu}.
\end{align}
The expression for Weyl scalar $\Psi_{4}$ is given by 

\begin{align}
\Psi_{4} = -\frac{1}{8}\Big((1+\zeta\bar{\zeta})^{2}\lambda_{\bar{\zeta}\bar{\zeta}}(\kappa\partial_{v}\lambda_{\zeta\zeta}+\partial^{2}_{v}\lambda_{\zeta\zeta})\Big)\rho^{2}+\mathcal{O}(\rho^{3}) .\label{psi4ext}
\end{align}
Similarly, one can compute the Weyl scalar $\Psi_{4}$ for extremal case, and get

\begin{align}
\Psi_{4} = -\frac{1}{8}\Big((1+\zeta\bar{\zeta})^{2}\lambda_{\bar{\zeta}\bar{\zeta}}\partial^{2}_{v}\lambda_{\zeta\zeta}\Big)\rho^{2}+\mathcal{O}(\rho^{3}). \label{psi4dextmn}
\end{align}

\subsection{Weyl Tensor for full metric}
For the full metric the expression for NP tetrads are quite large, however we  display here some Weyl tensor components that would be relevant for computing $\psi_4$.  
To the linear order in $\rho$ and $\kappa$ for the full metric (\ref{m}) with $\theta_{A}=\mathcal{C}(x^{A})$, the two shortest and relevant Weyl tensor components are
\begin{align}
   C_{\zeta v\zeta v} =& -\frac{\rho}{2}\partial^{2}_{v}\lambda_{\zeta\zeta}+\mathcal{O}(\rho^{2}) \\
   C_{\zeta v\bar{\zeta} v} =& \frac{\rho}{{6 (\zeta\bar{\zeta}+1)^2}}\Big(\kappa (2 (\zeta\bar{\zeta})^{2} \partial_{v}\lambda_{\zeta\bar{\zeta}}+4 \zeta\bar{\zeta} \partial_{v}\lambda_{\zeta\bar{\zeta}}+2 \partial_{v}\lambda_{\zeta\bar{\zeta}}-(\zeta\bar{\zeta})^{2} \partial_{\zeta}\theta_{\bar{\zeta}}- (\zeta\bar{\zeta}+1)^2 \partial_{\bar{\zeta}}\theta_{\zeta}- \nonumber \\
   & 2 \zeta\bar{\zeta} \partial_{\zeta}\theta_{\bar{\zeta}}-\partial_{\zeta}\theta_{\bar{\zeta}}+4)\Big)+\mathcal{O}(\rho^{2}).
\end{align}
From these components it is apparent that the NP scalar would depend on $\lambda$, $\theta$ and their derivatives.


\providecommand{\href}[2]{#2}\begingroup\raggedright\endgroup


\begin{thebibliography}{10}

\bibitem{PhysRevLett.116.061102}
{\scshape LIGO Scientific Collaboration and Virgo Collaboration} collaboration,
  \emph{Observation of gravitational waves from a binary black hole merger},
  \href{https://doi.org/10.1103/PhysRevLett.116.061102}{\emph{Phys. Rev. Lett.}
  {\bfseries 116} (2016) 061102}.

\bibitem{PhysRevLett.125.101102}
{\scshape LIGO Scientific Collaboration and Virgo Collaboration} collaboration,
  \emph{Gw190521: A binary black hole merger with a total mass of $150\text{
  }\text{ }{M}_{\ensuremath{\bigodot}}$},
  \href{https://doi.org/10.1103/PhysRevLett.125.101102}{\emph{Phys. Rev. Lett.}
  {\bfseries 125} (2020) 101102}.

\bibitem{Zeldovich:1974gvh}
Y.~B. Zel'dovich and A.~G. Polnarev, \emph{{Radiation of gravitational waves by
  a cluster of superdense stars}}, {\emph{Sov. Astron.} {\bfseries 18} (1974)
  17}.

\bibitem{Braginsky-Thorne}
V.~B. Braginsky and L.~P. Grishchuk, \emph{{Kinematic resonance and the memory
  effect in free mass gravitational antennas}}, {\emph{Sov. Phys. JETP}
  {\bfseries 62} (1985) 427}.

\bibitem{Braginsky}
V.~B. Braginsky and K.~S. Thorne, \emph{{Present status of gravitational-wave
  experiments}}, {\emph{In Proceedings of the ninth international conference on
  general relativity and gravitation (Jena, 1980), pages 239-253, Cambridge
  Univ. Press} (1983) }.

\bibitem{PhysRevLett.67.1486}
D.~Christodoulou, \emph{Nonlinear nature of gravitation and gravitational-wave
  experiments}, \href{https://doi.org/10.1103/PhysRevLett.67.1486}{\emph{Phys.
  Rev. Lett.} {\bfseries 67} (1991) 1486}.

\bibitem{Bieri:2013ada}
L.~Bieri and D.~Garfinkle, \emph{{Perturbative and gauge invariant treatment of
  gravitational wave memory}},
  \href{https://doi.org/10.1103/PhysRevD.89.084039}{\emph{Phys. Rev. D}
  {\bfseries 89} (2014) 084039}
  [\href{https://arxiv.org/abs/1312.6871}{{\ttfamily 1312.6871}}].

\bibitem{Bieri:2015yia}
L.~Bieri, D.~Garfinkle and S.-T. Yau, \emph{{Gravitational Waves and Their
  Memory in General Relativity}}, {\emph{The Centenary of General Relativity,
  volume 20 of Surveys in Differential Geometry} (2015) }
  [\href{https://arxiv.org/abs/1505.05213}{{\ttfamily 1505.05213}}].

\bibitem{PhysRevD.80.024002}
M.~Favata, \emph{Post-newtonian corrections to the gravitational-wave memory
  for quasicircular, inspiralling compact binaries},
  \href{https://doi.org/10.1103/PhysRevD.80.024002}{\emph{Phys. Rev. D}
  {\bfseries 80} (2009) 024002}.

\bibitem{Islo:2019qht}
K.~Islo, J.~Simon, S.~Burke-Spolaor and X.~Siemens, \emph{{Prospects for Memory
  Detection with Low-Frequency Gravitational Wave Detectors}}, {\emph{arXiv}
  (2019) } [\href{https://arxiv.org/abs/1906.11936}{{\ttfamily 1906.11936}}].

\bibitem{PhysRevLett.117.061102}
P.~D. Lasky, E.~Thrane, Y.~Levin, J.~Blackman and Y.~Chen, \emph{Detecting
  gravitational-wave memory with ligo: Implications of gw150914},
  \href{https://doi.org/10.1103/PhysRevLett.117.061102}{\emph{Phys. Rev. Lett.}
  {\bfseries 117} (2016) 061102}.

\bibitem{PhysRevD.101.083026}
O.~M. Boersma, D.~A. Nichols and P.~Schmidt, \emph{Forecasts for detecting the
  gravitational-wave memory effect with advanced ligo and virgo},
  \href{https://doi.org/10.1103/PhysRevD.101.083026}{\emph{Phys. Rev. D}
  {\bfseries 101} (2020) 083026}.

\bibitem{Bondi:1962px}
H.~Bondi, M.~van~der Burg and A.~Metzner, \emph{{Gravitational waves in general
  relativity. 7. Waves from axisymmetric isolated systems}},
  \href{https://doi.org/10.1098/rspa.1962.0161}{\emph{Proc. Roy. Soc. Lond. A}
  {\bfseries A269} (1962) 21}.

\bibitem{Strominger:2014pwa}
A.~Strominger and A.~Zhiboedov, \emph{{Gravitational Memory, BMS
  Supertranslations and Soft Theorems}},
  \href{https://doi.org/10.1007/JHEP01(2016)086}{\emph{JHEP} {\bfseries 01}
  (2016) 086} [\href{https://arxiv.org/abs/1411.5745}{{\ttfamily 1411.5745}}].

\bibitem{Strominger:2017zoo}
A.~Strominger, \emph{{Lectures on the Infrared Structure of Gravity and Gauge
  Theory}}. Princeon University Press, 3, 2017,
  [\href{https://arxiv.org/abs/1703.05448}{{\ttfamily 1703.05448}}].

\bibitem{PhysRevLett.116.231301}
S.~W. Hawking, M.~J. Perry and A.~Strominger, \emph{Soft hair on black holes},
  \href{https://doi.org/10.1103/PhysRevLett.116.231301}{\emph{Phys. Rev. Lett.}
  {\bfseries 116} (2016) 231301}.

\bibitem{Hawking:2016sgy}
S.~W. Hawking, M.~J. Perry and A.~Strominger, \emph{{Superrotation Charge and
  Supertranslation Hair on Black Holes}},
  \href{https://doi.org/10.1007/JHEP05(2017)161}{\emph{JHEP} {\bfseries 05}
  (2017) 161} [\href{https://arxiv.org/abs/1611.09175}{{\ttfamily
  1611.09175}}].

\bibitem{Zhang:2017geq}
P.-M. Zhang, C.~Duval, G.~Gibbons and P.~Horvathy, \emph{{Soft gravitons and
  the memory effect for plane gravitational waves}},
  \href{https://doi.org/10.1103/PhysRevD.96.064013}{\emph{Phys. Rev. D}
  {\bfseries 96} (2017) 064013}
  [\href{https://arxiv.org/abs/1705.01378}{{\ttfamily 1705.01378}}].

\bibitem{Pasterski:2015tva}
S.~Pasterski, A.~Strominger and A.~Zhiboedov, \emph{{New Gravitational
  Memories}}, \href{https://doi.org/10.1007/JHEP12(2016)053}{\emph{JHEP}
  {\bfseries 12} (2016) 053}
  [\href{https://arxiv.org/abs/1502.06120}{{\ttfamily 1502.06120}}].

\bibitem{hawking2015information}
S.~W. Hawking, \emph{The information paradox for black holes},  2015.

\bibitem{PhysRevLett.105.111103}
G.~Barnich and C.~Troessaert, \emph{Symmetries of asymptotically flat
  four-dimensional spacetimes at null infinity revisited},
  \href{https://doi.org/10.1103/PhysRevLett.105.111103}{\emph{Phys. Rev. Lett.}
  {\bfseries 105} (2010) 111103}.

\bibitem{Barnich:2011ct}
G.~Barnich and C.~Troessaert, \emph{{Supertranslations call for
  superrotations}}, \href{https://doi.org/10.22323/1.127.0010}{\emph{PoS}
  {\bfseries CNCFG2010} (2010) 010}
  [\href{https://arxiv.org/abs/1102.4632}{{\ttfamily 1102.4632}}].

\bibitem{PhysRevLett.116.091101}
L.~Donnay, G.~Giribet, H.~A. Gonz\'alez and M.~Pino, \emph{Supertranslations
  and superrotations at the black hole horizon},
  \href{https://doi.org/10.1103/PhysRevLett.116.091101}{\emph{Phys. Rev. Lett.}
  {\bfseries 116} (2016) 091101}.

\bibitem{Donnay2016}
L.~Donnay, G.~Giribet, H.~A. Gonz{\'a}lez and M.~Pino, \emph{Extended
  symmetries at the black hole horizon},
  \href{https://doi.org/10.1007/JHEP09(2016)100}{\emph{Journal of High Energy
  Physics} {\bfseries 2016} (2016) 100}.

\bibitem{PhysRevD.98.124016}
L.~Donnay, G.~Giribet, H.~A. Gonz\'alez and A.~Puhm, \emph{Black hole memory
  effect}, \href{https://doi.org/10.1103/PhysRevD.98.124016}{\emph{Phys. Rev.
  D} {\bfseries 98} (2018) 124016}.

\bibitem{Blau:2015nee}
M.~Blau and M.~O'Loughlin, \emph{{Horizon Shells and BMS-like Soldering
  Transformations}}, \href{https://doi.org/10.1007/JHEP03(2016)029}{\emph{JHEP}
  {\bfseries 03} (2016) 029}
  [\href{https://arxiv.org/abs/1512.02858}{{\ttfamily 1512.02858}}].

\bibitem{Bhattacharjee}
S.~Bhattacharjee and A.~Bhattacharyya, \emph{Soldering freedom and
  bondi-metzner-sachs-like transformations},
  \href{https://doi.org/10.1103/PhysRevD.98.104009}{\emph{Phys. Rev. D}
  {\bfseries 98} (2018) 104009}.

\bibitem{OLoughlin:2018ebk}
M.~O'Loughlin and H.~Demirchian, \emph{{Geodesic congruences, impulsive
  gravitational waves and gravitational memory}},
  \href{https://doi.org/10.1103/PhysRevD.99.024031}{\emph{Phys. Rev. D}
  {\bfseries 99} (2019) 024031}
  [\href{https://arxiv.org/abs/1808.04886}{{\ttfamily 1808.04886}}].

\bibitem{PhysRevD.100.084010}
S.~Bhattacharjee, S.~Kumar and A.~Bhattacharyya, \emph{Memory effect and
  bms-like symmetries for impulsive gravitational waves},
  \href{https://doi.org/10.1103/PhysRevD.100.084010}{\emph{Phys. Rev. D}
  {\bfseries 100} (2019) 084010}.

\bibitem{PhysRevD.102.044041}
S.~Bhattacharjee and S.~Kumar, \emph{Memory effect and bms symmetries for
  extreme black holes},
  \href{https://doi.org/10.1103/PhysRevD.102.044041}{\emph{Phys. Rev. D}
  {\bfseries 102} (2020) 044041}.

\bibitem{p7}
I.~Chakraborty and S.~Kar, \emph{Geodesic congruences in exact plane wave
  spacetimes and the memory effect},
  \href{https://doi.org/10.1103/PhysRevD.101.064022}{\emph{Phys. Rev. D}
  {\bfseries 101} (2020) 064022}.

\bibitem{Kolekar:2017yoi}
S.~Kolekar and J.~Louko, \emph{{Gravitational memory for uniformly accelerated
  observers}}, \href{https://doi.org/10.1103/PhysRevD.96.024054}{\emph{Phys.
  Rev. D} {\bfseries 96} (2017) 024054}
  [\href{https://arxiv.org/abs/1703.10619}{{\ttfamily 1703.10619}}].

\bibitem{Compere:2019rof}
G.~Comp\`ere, J.~Long and M.~Riegler, \emph{{Invariance of Unruh and Hawking
  radiation under matter-induced supertranslations}},
  \href{https://doi.org/10.1007/JHEP05(2019)053}{\emph{JHEP} {\bfseries 05}
  (2019) 053} [\href{https://arxiv.org/abs/1903.01812}{{\ttfamily
  1903.01812}}].

\bibitem{Strominger2017}
A.~Strominger and A.~Zhiboedov, \emph{Superrotations and black hole pair
  creation}, \href{https://doi.org/10.1088/1361-6382/aa5b5f}{\emph{Classical
  and Quantum Gravity} {\bfseries 34} (2017) 064002}.

\bibitem{PhysRevD.89.064008}
A.~Tolish and R.~M. Wald, \emph{Retarded fields of null particles and the
  memory effect}, \href{https://doi.org/10.1103/PhysRevD.89.064008}{\emph{Phys.
  Rev. D} {\bfseries 89} (2014) 064008}.

\bibitem{PhysRevD.45.520}
K.~S. Thorne, \emph{Gravitational-wave bursts with memory: The christodoulou
  effect}, \href{https://doi.org/10.1103/PhysRevD.45.520}{\emph{Phys. Rev. D}
  {\bfseries 45} (1992) 520}.

\bibitem{PhysRevD.90.044060}
A.~Tolish, L.~Bieri, D.~Garfinkle and R.~M. Wald, \emph{Examination of a simple
  example of gravitational wave memory},
  \href{https://doi.org/10.1103/PhysRevD.90.044060}{\emph{Phys. Rev. D}
  {\bfseries 90} (2014) 044060}.

\bibitem{Chandrasekaran:2018aop}
V.~Chandrasekaran, E.~E. Flanagan and K.~Prabhu, \emph{{Symmetries and charges
  of general relativity at null boundaries}},
  \href{https://doi.org/10.1007/JHEP11(2018)125}{\emph{JHEP} {\bfseries 11}
  (2018) 125} [\href{https://arxiv.org/abs/1807.11499}{{\ttfamily
  1807.11499}}].
  
  \bibitem{Fernandes:2020jto}
    K.~Fernandes, D.~Ghosh and A.~Virmani,
\emph{Horizon Hair from Inversion Symmetry,''}
[arXiv:2008.04365 [hep-th]].

\bibitem{Khera:2020mcz}
N.~Khera, B.~Krishnan, A.~Ashtekar and T.~De~Lorenzo, \emph{{Inferring the
  gravitational wave memory for binary coalescence events}},
  \href{https://arxiv.org/abs/2009.06351}{{\ttfamily 2009.06351}}.
  


\end{thebibliography}
\end{document}